\documentclass[conference]{IEEEtran}
\IEEEoverridecommandlockouts

\normalsize
\usepackage{cite}
\usepackage[normalem]{ulem}
\useunder{\uline}{\ul}{}
\ifCLASSINFOpdf
   \usepackage[pdftex]{graphicx}
   \DeclareGraphicsExtensions{.pdf}
\else
   \usepackage[dvips]{graphicx}
   \DeclareGraphicsExtensions{.eps}
\fi

 \ifCLASSOPTIONcompsoc
	\usepackage[caption=false,font=footnotesize,labelfont=sf,textfont=sf]{subfig}
\else
	\usepackage[caption=false,font=footnotesize]{subfig}
\fi

\RequirePackage{lineno}

\usepackage{color}
\usepackage[usenames]{xcolor}
\usepackage{microtype}

\usepackage{amsmath,amssymb,amsfonts,bbm}
\usepackage{booktabs}
\usepackage[inline]{enumitem}
\usepackage{graphicx, float}
\usepackage{nccmath}
\usepackage{multirow}

\usepackage[backgroundcolor=cyan,bordercolor=gray,textsize=small,textwidth=5.8cm]{todonotes}

\usepackage{stfloats}

\usepackage{url}

 \DeclareMathOperator{\diag}{diag}

\usepackage[ruled,vlined,linesnumbered]{algorithm2e}
\usepackage{algpseudocode}
\usepackage[flushleft]{threeparttable}

\makeatletter
\def\BState{\State\hskip-\ALG@thistlm}
\makeatother

\usepackage[backgroundcolor=cyan,bordercolor=gray,textsize=small,textwidth=5.8cm]{todonotes}

\DeclareMathOperator*{\trace}{Tr}

\DeclareMathOperator*{\rank}{rank}

\makeatletter
\def\BState{\State\hskip-\ALG@thistlm}
\makeatother

\usepackage{textcomp}

\newtheorem{remark}{Remark}

\def\BibTeX{{\rm B\kern-.05em{\sc i\kern-.025em b}\kern-.08em T\kern-.1667em\lower.7ex\hbox{E}\kern-.125emX}}

\newcommand{\vertiii}[1]{{\left\vert\kern-0.25ex\left\vert\kern-0.25ex\left\vert #1
 \right\vert\kern-0.25ex\right\vert\kern-0.25ex\right\vert}}

\newcommand{\vect}[1]{\boldsymbol{\mathbf{#1}}}

\newlist{myitemize}{itemize}{3}
\setlist[myitemize]{noitemsep, topsep=0pt}
\setlist[myitemize,1]{label=\protect\itemcircle ,leftmargin=1.0in}
\setlist[myitemize,2]{label=$\rightarrow$,leftmargin=1em}
\setlist[myitemize,3]{label=$\diamond$}

\usepackage{tcolorbox}
\definecolor{BlueGray}{HTML}{546E7A}
\definecolor{BlueGrayLight}{HTML}{607D8B}
\definecolor{Amber}{HTML}{FF8F00}
\definecolor{Orange}{HTML}{FB8C00}
\definecolor{DeepOrange}{HTML}{FF7043}
\definecolor{Indigo}{HTML}{3949AB}
\definecolor{Teal}{HTML}{00897B}
\definecolor{LightGreen}{HTML}{7CB342}
\definecolor{LightBlue}{HTML}{0277BD}
\definecolor{Purple}{HTML}{AB47BC}
\definecolor{DeepPurple}{HTML}{7E57C2}

\begin{document}

%Multi-User Beam Scheduling with Deep Reinforcement Learning in Integrated Sensing and Communication Systems 
\title{
Beam Scheduling for Cross-Layer ISAC: \\A Deep Reinforcement Learning  Approach\\ 
% Cross-Layer Beam Scheduling for Integrated Sensing and Communication via Reinforcement Learning\\

\thanks{X. Wang and H. V. Cheng are with the Department of Electrical and Computer Engineering, Aarhus University, Denmark (email: xiyuw@ece.au.dk, hvc@ece.au.dk). 

G. Berardinelli, P. Popovski, and R. Adeogun are with the Department of Electronic Systems, Aalborg University, Denmark (email: ra@es.aau.dk, gb@es.aau.dk,  petarp@es.aau.dk).

This work is presented in part at EuCNC/6G Summit 2025. This research is supported in part by the HORIZON JU-SNS-2022-STREAM-B-01-02 CENTRIC project (Grant Agreement No.101096379). Xiyu Wang is also supported in part by the European Union’s Horizon 2020 research and innovation programme under the Marie Sk\l{}odowska-Curie Grant agreement No. 101146247. Hei Victor Cheng's work is supported in part by the Aarhus Universitets Forskningsfond project No. AUFF39001 and NordForsk Nordic University Cooperation on Edge Intelligence (Grant No. 168043).
}
}\author{\IEEEauthorblockN{Xiyu Wang, Gilberto Berardinelli, Hei Victor Cheng,  Petar Popovski, and Ramoni Adeogun}
}

\maketitle

\begin{abstract}
% The  paradigm enables diverse applications via merging communication and sensing modules. 

Resource allocation in integrated sensing and communication (ISAC) systems needs
to be optimized to balance the requirements of the communication and sensing
modules considering complicated cross-layer data traffic and queue status in
dynamic multi-user environments.
This paper studies the beam allocation for cross-layer ISAC that achieves low-latency communication and minimizes sensing parameters estimation error. To handle the complex coupling between practical data buffer dynamics and varying wireless channels,  we propose a deep reinforcement learning (DRL)-assisted approach. 
% The proposed method allocates beams to jointly minimize delay and sensing errors by implicitly capturing the intricate interplay between buffers and channel conditions. 
Rather than relying on explicit channel state information, the DRL-assisted beam allocation reduces feedback overhead by leveraging sensing observations. Simulation results verify that the DRL framework effectively takes buffer status into account and adapts to the wireless environment while allocating resources. The proposed multi-beam scheme improves overall throughput with only modest delay increases. Finally, the DRL-assisted beam management achieves both communication and sensing performance close to that of the genie-aided benchmark with perfect angle-of-departure (AoD) knowledge. These contributions advance the state-of-the-art intelligent resource management for ISAC systems.   
\end{abstract}

\section{Introduction}

Sensing is gaining increasing importance in sixth-generation (6G) communications, as it provides richer information to aid communication. Integrated sensing and communication (ISAC) has become a crucial component of 6G technologies for improving quality of service and expanding application scenarios. In ISAC, the sensing module retrieves information such as locations and velocities of objects of interest through parameter estimation from received radio frequency (RF) signals; the communication module focuses on reliable, low-latency data transmission to users over wireless channels. The two integrated modules share hardware and RF resources, including time, spectrum, energy, and beam space.
Managing resources allocated to both modules to fulfill their respective requirements is nontrivial in ISAC, especially considering cross-layer requirements in dynamic multi-user scenarios \cite{liu_integrated_2022}.

Owing to the maturation of millimeter-wave (mmWave) space division multiplexing, resource allocation in the beam space is promising~\cite{xue_survey_2024}. mmWave signals suffer from severe free-space path loss and exhibit sparse channel properties, which necessitate directive beamforming to maintain an adequate link budget. Equipping base stations (BSs) with large antenna arrays generates narrow, high-gain beams that compensate for the path loss while simultaneously providing the angular resolution for accurate sensing parameter estimation. This angular sparsity makes beam space a natural and efficient resource dimension, offering an additional degree of freedom to the time and frequency domains \cite{zhuo_2024_multibeam_isac}.

That said, resource allocation within beam space is non-trivial in practice. Narrow beams in the mmWave spectrum are susceptible to beam drifting, i.e., the beamforming target is not aligned with the center of the beam, especially in high-mobility environments \cite{Zhang2021BeamDrift}. In dynamic environments, beams should be predicted to prevent beam failure and avoid losing track of targets. Wide beams cover a large field of view (FoV) but sacrifice signal strength concentration. 
Beyond physical-layer beam alignment, practical ISAC deployments also need to accommodate stringent quality-of-service (QoS) demands arising from upper-layer data traffic. Conventional single-layer beam management yields poor performance in fulfilling communication latency, sensing accuracy, and QoS requirements simultaneously. The complex coupling between communication and sensing objectives, together with environmental dynamics, makes traditional methods difficult to apply. However, this situation can be turned into an advantage by noting that
sensing provides informative observations of the environment and those can be exploited to assist resource management. Motivated by this, this paper develops a deep reinforcement learning (DRL) framework that relies upon the sensing module to facilitate cross-layer interaction between the queue management and physical-layer beam management for ISAC.

\subsection{Related Work}
Various frameworks have been proposed for aligning beams and resolving beam drifting. 3GPP standards adopt periodic beam sweeping as the solution \cite{xue_survey_2024}. The sweeping frequency should be optimized according to system kinematics. The work \cite{Zhang2021BeamDrift} proposes a varying-beamwidth transmission to mitigate beam drifting in a single user scenario. This protocol relies on user feedback, which can yield high overhead and result in late beam updates.
Sensing helps characterize the wireless environment and can potentially predict beam allocation with reduced tracking overhead \cite{meng_vehicular_2023}.
Available works \cite{liu_integrated_2022, MugenPeng2023, chen_multiuser_2023, liu_radar-assisted_2020} deploy a sensing receiver to capture reflected echoes, from which the AoD is estimated and predicted to improve beam alignment using state evolution models.
Obtaining prior knowledge of state evolution is challenging, especially in multi-user dynamic environments. 

ISAC systems generate a large volume of information on data traffic and sensing requirements that need to be processed at control centers. In this case, BS resource allocation and scheduling significantly affect requirement fulfillment and its efficiency. Available work primarily investigates waveform design and resource management \cite{liu_radar-assisted_2020, deep_rl_liu_2024_wcnc, Li_drl_bm_2022, chen_multiuser_2023, Chen2024, smida_inband_2024} by optimizing the metrics of the physical layer or media access control (MAC) layer separately \cite{meng_cooperative_2025}. 
% However, conventional single-layer resource management yields poor performance when simultaneously accommodating a high volume of data transmission, accurate parameter estimation, and stringent quality of service (QoS) demands. 
Only a few attempts have been made to characterize upper-layer performance that account for queuing effects. \cite{xie_optimal_2023} investigates a power and time allocation scheme that jointly minimizes queuing delay and maximizes target detection probability in a single-input single-output (SISO) system. 
The work \cite{xie_noma_isac_allocation_2025} extends the allocation strategy to MIMO ISAC systems. Additionally, \cite{ding_joint_2022} minimizes the transmit power while guaranteeing radar SINR and delay constraints. The stochastic scheduling schemes proposed in \cite{xie_optimal_2023, xie_noma_isac_allocation_2025} assume known quantized channel state information (CSI). The scheduling also requires the calculation of state transition probabilities, whose complexity grows exponentially with the quantization level of the channel and packet arrival rates. This scalability issue presents significant challenges for real-time implementation in dynamic environments.

Learning-based algorithms stand out because they can capture complex dynamic environments to optimize beam management. Beam selection has been done using deep neural networks to maximize the sum rate \cite{Chen2024}. A deep reinforcement learning (DRL) approach to optimize the beam and power allocation in an ISAC system is studied in \cite{deep_rl_liu_2024_wcnc}. It maximizes the sum rate while taking estimation precision, i.e., Cramér-Rao lower bound (CRLB), as a penalty. However, resource allocation is optimized for a single time instant, and the proposed method requires genie information for the CRLB calculation. Management of beamwidth and interference between a communication train and a sensing target in a fast-moving railway scenario is studied in \cite{Li_drl_bm_2022}. The proposed DRL model requires user signal-to-interference-plus-noise ratio (SINR) feedback as observations, which increases the scheduling overhead. Moreover, these learning-based algorithms focus on physical-layer metrics.

\subsection{Motivation and Contribution}

This paper studies beam allocation for dynamic multi-user cross-layer ISAC systems, where the inter-dependencies between queue dynamics, channel variations, and dual sensing-communication objectives create a significant design challenge.  
A few approaches to resource management for cross-layer ISAC systems have shown promise; however, they often rely on CSI to achieve optimal performance. 
% For instance, \cite{xie_noma_isac_allocation_2025} proposes a stochastic scheduling scheme with a closed-form solution for the rate adaptation given quantized CSI. 
Acquiring timely and accurate CSI introduces substantial control signaling overhead that can degrade system efficiency. In practical ISAC deployments, this is increasingly challenging due to the dynamic nature of wireless environments and the dual sensing-communication functionality. Moreover, even with perfect CSI, integrating buffer state information into analytical problem formulations introduces additional complexity, as the inter-dependencies between queue dynamics, channel conditions, and beam allocation decisions are difficult to capture in tractable closed-form expressions. 

These gaps motivate a DRL-based framework that jointly optimizes communication latency and sensing accuracy across layers, eliminates explicit CSI feedback by utilizing sensing observations, and implicitly captures the intractable analytical model of buffer, channel, and beam interactions through policy learning. 
% The design learns to allocate beams by directly observing system states--including buffer states and reflected echoes--without requiring explicit analytical models. The framework adapts to varying data arrivals and time-varying propagation channels. In addition, it shortens communication latency and requires minimal user feedback by leveraging the sensing capability of the ISAC scheme. 
This work extends \cite{wang_multiuser_beamforming_2025}, which demonstrated that adopting multiple beams is an effective strategy for preventing beam drifting for mobile users. While \cite{wang_multiuser_beamforming_2025} focuses solely on physical-layer communication performance aided by sensing in a narrowband system, this paper incorporates cross-layer queue dynamics, adopts a wideband orthogonal frequency division multiplexing (OFDM) framework, and integrates range and velocity estimation into the sensing module.  

Our main contributions are as follows.
\begin{itemize}
    \item We formulate the beam allocation problem for joint optimization of communication latency and sensing parameter estimation in dynamic multi-user cross-layer ISAC systems. We show that the complexity of incorporating buffer, channel, and communication-sensing objectives into a tractable expression hinders traditional optimization methods.  
    \item 
    We propose the DRL-assisted beam allocation approach that takes sensing-derived observations, i.e., the beamforming output of the reflected echo, as the main state, thereby eliminating the need for explicit CSI feedback. The approach learns beam scheduling policies that capture the complex coupling between buffer, channel, and communication-sensing objectives without requiring explicit channel states and transition probability matrices. 
    \item Extensive experiments are conducted to demonstrate that the proposed DRL-assisted method: (i) adapts to the dynamics of wireless channels and data traffic patterns; (ii) distinguishes between clutters from user targets; and (iii) achieves communication latency and sensing performance close to the genie-aided benchmark with perfect AoD information, while requiring minimal user feedback.
\end{itemize}

% The contributions are the following. 
% A heuristic AoD-based method that uses approximated CRLB of AoD for grouping users and allocating beams is first developed.
% We then propose a DRL-assisted approach for user scheduling and beam allocation. The DRL model takes the beamforming output of the reflected echo as states and optimizes the policy that maximizes packet throughput. 
% Comparative results in random mobile scenarios show that the DRL-assisted beam management outperforms the periodical beam sweeping and the heuristic AoD-assisted method, and is robust to the user movements and speed. Therefore, the proposed methods adjust beams timely and mitigate beam drifting. Furthermore, our approach verifies that sensing helps simplify beam management without needing prior information or user feedback.   

\begin{figure}
    \centering
    \includegraphics[width=0.8\linewidth]{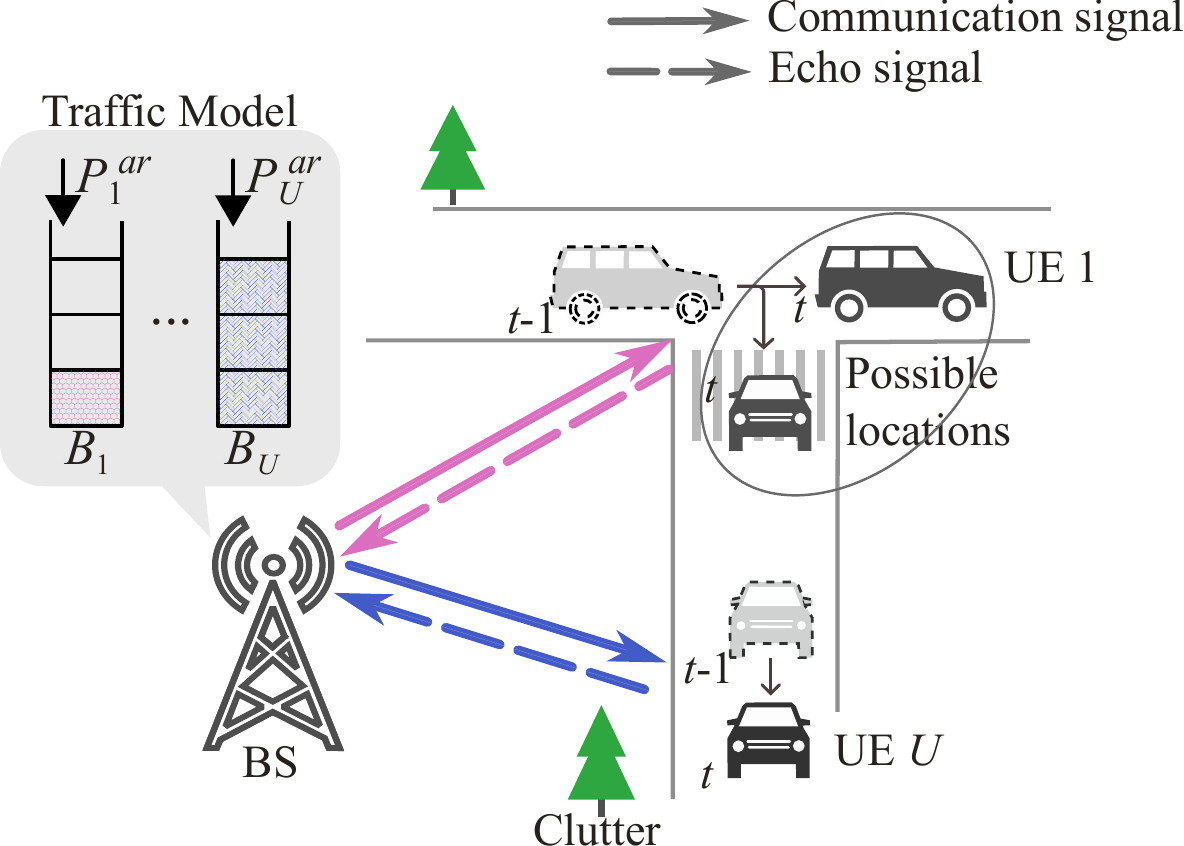}
    \caption{A multi-user mono-static ISAC system}
    \label{fig:system_model}
\end{figure}

\subsection{Notations and Paper Structures}
Throughout the paper, scalars are denoted by italic font letters $a$, vectors and matrices are represented by 
lower-case $\vect{a}$ and upper-case $\vect{A}$ boldface letters, respectively. Complex scalars 
are assumed, and their set is denoted by $\mathbb{C}$. The Euclidean norm of a vector $\vect{a}$ is represented 
by $\|\vect{a}\|$, the zero norm is represented by $\|\vect{a}\|_0$ and the absolute value of $a$ is represented by $|a|$. The $n\times n$ identity matrix is 
$\vect{I}_n$, and the subscript $n$ may be omitted sometimes for simplicity. The conjugate-transpose, 
conjugate, 
and transpose of a matrix $\vect{A}$ are $\vect{A}^H$, $\vect{A}^*$ and $\vect{A}^T$, 
respectively. The trace, the rank, and the determinant of a matrix $\vect{A}$ are represented by 
$\trace(\vect{A})$, $\rank (\vect{A})$ and $|\vect{A}|$. The $n$-th column of matrix $\vect{A}$ is represented as $[\vect{A}]_n$. $\mathcal{CN}(\vect{a}, 
\vect{A})$ denotes circularly symmetric complex Gaussian vector with mean 
$\vect{a}$ and covariance matrix $\vect{A}$, and $\mathcal{N}(\cdot)$ denotes the real Gaussian distribution. The indicator function is $\mathbbm{1}$.  The statistical expectation is 
$\mathrm{E}\{\cdot\}$. The imaginary unit is $j=\sqrt{-1}$.

The remainder of this paper is organized as follows. The system model is introduced in Section \ref{sec:system_model}, and the communication and the sensing metrics are presented in Section \ref{sec:isac_req}. The beam management problem of the ISAC system is introduced in Section \ref{sec:objective}, where a heuristic AoD-based method is proposed. Section \ref{sec:drl_method} introduces the proposed DRL-assisted approach. Section \ref{sec:simulation} presents the numerical results and the discussion. Section \ref{sec:conclusion} concludes the paper.

\section{System Model}

\label{sec:system_model}
Consider a multi-user mono-static ISAC system comprising one BS and $U$ users, as illustrated in Fig.~\ref{fig:system_model}. 
Assuming downlink, all users are communication users awaiting data transmission from the BS. Sensing is performed opportunistically in this communication-optimized setting and shares resources with the communication module. Moreover, there exists $C$ stationary clutters in the environment, whose locations are unknown to the transceiver.

The BS is dual-functional, and is equipped with an $N_T$-antenna uniform linear array (ULA) transmitter and an $N_R$-antenna ULA receiver for simultaneous data transmission and sensing parameters estimation.  We assume that the BS with co-located transmitter and receiver works in full-duplex mode, and antenna isolation between the transmitter and receiver arrays has reduced the self-interference to a negligible level, and thus the residual self-interference is ignored in the sensing model. Each user is assumed to be equipped with a single antenna. This allows us to focus on the beam management at the BS. The work can be extended to multi-antenna users at a cost of increased complexity by incorporating user beamforming into the framework.
% Without loss of generality, we assume that the number of users that the BS can serve is smaller than the number of radio frequency (RF) chains of the BS transmitter array. That is, the BS can simultaneously serve $U\leq N_T$ users in the same time-frequency resource.
In a given transmission time interval (TTI), we assume that the number of simultaneously served user does not exceed the number of RF chains, that is $U\le N_T$. It is important to note that this constraint only applies to the number of scheduled users within a TTI, rather than the total number of users in the system. In practical deployments, a BS can support a much larger number of users through MAC layer scheduling that selects a subset of users to serve at each TTI. Such an extension to large-scale system is left for future work.
In this work, we assume that the BS selects beams from a discrete Fourier transform (DFT)-based codebook, which is widely adopted for its implementation simplicity, as in LTE systems \cite{yang_dft-based_2010}. 

\subsection{Traffic Model} \label{sec:traffic_model}
For the downlink communication, the BS maintains $U$ buffers, one for each user.  Each buffer size is finite and may differ between users. The buffer size for the $u$-th user is denoted as $B_u$. The BS delivers packets in a first-come, first-served manner. Packets for each user arrive stochastically at the BS, following an independent and identically distributed (i.i.d.) Bernoulli distribution with arrival probabilities denoted as $P_{1}^{ar}, \cdots, P_{U}^{ar}$, respectively for each user. Due to finite buffer sizes, the BS faces a risk of overflow if accumulated packets cannot be transmitted promptly. Denote the buffer occupancy for all users at time instant $n$ as $q_1^{}[n], \cdots, q_U^{}[n]$. Newly arriving packets are dropped when the buffer is full, i.e., $q_u^{}[n] = B_u$. 

To accommodate practical scenarios with varying data-intensity requirements, user-specific latency constraints, also known as deadlines, are denoted by $D_1, \cdots, D_U$. The latency constraints are expressed in terms of the number of TTIs. When a data packet arrives in the $u$-th buffer, its latency is initialized as 0. Upon a transmission failure, the latency is incremented by 1, and the BS attempts retransmission provided that the packet has not expired. Let $\delta_u[n]$ denote the longest waiting time that the packets have in the $u$-th buffer at the $n$-th time instant. Owing to the first-come, first-served property, $\delta_u[n]$ quantifies the waiting time of the head-of-line packet, i.e., the next packet the BS attempts to transmit. To characterize the effect of the deadlines, the head-of-line packet in the $u$-th buffer with $\delta_u[n] = D_u$ TTIs is dropped due to deadline violations.  

To facilitate cross-layer analysis, we adopt a simplified packet-level abstraction where time is discretized into TTIs. Each packet occupies one TTI for transmission and thus, the transmission delay is modeled as a constant, equal to one TTI. This assumption allows us to focus primarily on queuing delay, which arises from the traffic model. 

\subsection{Transmitted OFDM Waveform}
The BS adopts an OFDM waveform for communication. We assume the channel is block-fading, that is, the channels are frequency-selective, and they stay the same within one OFDM frame, and experience independent small-scale fading across OFDM frames. A TTI spans $M$ OFDM frames. Considering the limited user velocities, a user remains within the coverage of a single beam over one TTI. As such, beam allocation is carried out every TTI. Nevertheless, in what follows, the signal representation is within one OFDM frame, and thus, the index of the frame is omitted to simplify notations until further notice.

The modulated symbol vector for the $U$ users at the $i$-th subcarrier and $l$-th OFDM symbol in the frequency domain is $\vect{s}_i[l] = [s^1_i[l], \cdots, s^U_i[l]] \in \mathbb{C}^{U\times 1}$, where $i = 0, \cdots, N_c-1$, $l = 0, \cdots, N_s-1$, and $N_c$ and $N_s$ are the number of subcarriers and the number of symbols in a frame, respectively. The symbols are drawn from a fixed constellation, and all users employ the same modulation scheme. Denote $P^1_i[l], \cdots, P^U_i[l]$ as the power allocated to the users at the $i$-th subcarrier and $l$-th symbol. We have $\vect{s}^u_i[l] (\vect{s}_i^u[l])^* = P_i^u[l]$, and $\mathrm{E}\{\vect{s}^u_i[l] (\vect{s}_i^q[l])^*\} = 0, q\neq u$.  And for any $i' \neq i$ or $l' \neq l$, $\mathrm{E}\{\vect{s}_i[l] \vect{s}^H_{i'}[l']\} = \vect{0}$.
The modulated symbol vector is then multiplied by the transmit beamforming vectors, followed by a linear combination to form the digital-precoded baseband signal, which is
\begin{equation}
    \vect{x}_i[l] = \Tilde{\vect{F}}\vect{V}\vect{s}_i[l]  
    \label{eq:tx_sig_1}
\end{equation}
where $\Tilde{\vect{F}} \in \mathbb{R}^{N_T \times N_T}$ is the DFT codebook matrix, and the $n_a$-th codeword in the DFT codebook is given by
\begin{equation}
    [\tilde{\vect{F}}]_{:,n_a} = \frac{1}{\sqrt{N_T}}\left[1, e^{j \omega(n_a)}, \cdots, e^{j(N_T-1)\omega(n_a)}\right],  
\label{eq:codebook}
\end{equation}
where $\omega(n_a) = \pi\frac{2n_a-1-N_T}{N_T}$ and $n_a = 1, \cdots, N_T$ is the index of angular bin.  
We denote $\vect{V} = [\vect{v}_1, \cdots, \vect{v}_U]\in \mathbb{R}^{N_T \times U}$ as the beam selection matrix.  We assume at most 3 beams are assigned to a user, which can be represented using the $\ell^0$ norm constraints on the columns of $\vect{V}$ as $\|\vect{v}_u\|_0 \le 3$. 
%In this case, the Frobenius Norm of $\vect{V}$ has the constraint $\|\vect{V}\|_F^2 \leq 3U$.  
One beam cannot be reused for multiple users such that $\vect{V}^T \vect{V}$ is a diagonal matrix\footnote{The strategy for preventing reusing beams will be described in Section \ref{sec:drl_method}.}. 

An $N_c$-point inverse discrete Fourier transform (IDFT) is carried out on a collection of $N_c$ digital-precoded baseband signal samples to convert the signal into the time domain. Then, a cyclic prefix (CP) is added to prevent inter-symbol interference (ISI). The length of CP, $T_{cp}$, is chosen to exceed the maximum of the channel impulse response and the round-trip delay between the BS and the furthest user. After conversion to analog, the transmitted time domain communication signal reads as
\begin{equation}
    \hat{\vect{x}}(t) = \frac{1}{\sqrt{N_c}} \sum_{l = 0}^{N_s -1} \sum_{i=0}^{N_c -1 } \vect{x}_i[l] e^{j 2\pi \Delta f i t } \mathrm{rect}\left(\frac{t-l T_s}{T_s}\right)
\end{equation}
where $T_s=1/\Delta f + T_{cp}$ is the symbol duration and $\Delta f$ is subcarrier spacing, and $\mathrm{rect}(t)$ is a rectangular pulse that takes the value one for $t\in[0,1]$ and zero otherwise. 

The up-converted time-domain signal $\hat{\vect{x}}(t)$ is received by the $U$ users. Meanwhile, the signal is reflected by them, and the echo signal is received by the BS receiver. These received signals in the time domain are first down-converted, followed by a sampling at the frequency of $N_c \Delta f$. Then, the time-domain signal is converted to the frequency domain using DFT. A detailed signal processing procedure can be found in, e.g., \cite{xiao_novel_2024, li_mimo-ofdm_2024}. It is assumed that a suitable parametrization is adopted to avoid inter-carrier interference and ISI. From now on, we assume the aforementioned signal processing procedure has been conducted.

% \question{Beamforming} 
% The transmitted signal model is the multi-stream beamforming, that is, the transmitted signal includes $U$ streams of OFDM symbols. This looks like the digital precoding to me. However, normally, digital precoding is frequency-selective. But here we assumed frequency-flat, as we only considered LoS channel. Do we need to specifically clarify which precoding scheme? 

\section{ISAC Communication and Sensing Tasks}
\label{sec:isac_req}
\subsection{Communication Performance Metric}
At the $u$-th user, the received frequency-domain signal of the $i$-th subcarrier and $l$-th symbol is written as
\begin{equation}
    y_i^u[l] = \vect{h}^u_i[l]^H \vect{x}_i[l] + w_i[l] ,
\end{equation}
where $w_i[l]\sim \mathcal{CN}(0,\sigma_{\omega}^2)$ is additive white Gaussian noise (AWGN) following i.i.d. zero-mean Gaussian distribution with noise power $\sigma_{\omega}^2$ across OFDM symbols and subcarriers,  and $\vect{h}^u_i[l]$ is the frequency channel response between the BS and $u$-th user. The channel is assumed Rician with Rician K-factor denoted as $K$ \cite{Va2017}, i.e.,
\begin{equation}
    \vect{h}^u_i[l] = \sqrt{\frac{K}{1+K}} \vect{h}^u_{i,\text{LoS}}[l] + \sqrt{\frac{1}{1+K}} \vect{h}^u_{i,\text{NLoS}}[l] 
\end{equation}
where the line-of-sight (LoS)  component is given by
\begin{equation}
    \vect{h}^u_{i,\text{LoS}}[l] =  \alpha (d_{u}) e^{j2\pi i \Delta f d_{u}/c} e^{j 2 \pi l T_s v_{u} f_c/c} \vect{a}(\phi_{u}) ,
\end{equation}
which is characterized by the delay $d_{u}/c$, Doppler shift $v_{u} f_c/c$, complex channel gain $\alpha (d_{u})$ and AoD $\phi_{u}$, where $c$ is the speed of light, $d_u$ is the distance between the user and the BS, $v_u$ is the user speed, and $f_c$ denotes the carrier frequency. Vector $\vect{a}(\theta) = \frac{1}{\sqrt{N_T}} [1, e^{j \theta}, \cdots, e^{j (N_T-1)\theta}]^T$ is the transmit steering vector, and $\theta = \pi \sin \Tilde{\theta}$ is normalized from physical AoD $\Tilde{\theta}$. 

The SINR of the user signal is computed as
\begin{equation}
    \gamma_{il}^u = \frac{ \| \vect{h}_i^u[l]^H \Tilde{\vect{F}} \vect{v}_u s_i^u[l] \|^2 }{\sum_{q=1,q\neq u}^{U}\|\vect{h}_i^u[l]^H \Tilde{\vect{F}} \vect{v}_q s_i^q[l] \|^2 +\sigma^2_{\omega} } ,
\label{eq:sinr_user_signal}
\end{equation}
\textcolor{black}{which depends on the beam and power allocation.}
Then, the communication rate \textcolor{black}{of the $m$-th OFDM symbol} can be calculated as
\begin{equation}
    \mathrm{R}_{u,m} = \sum_{l=0}^{N_s - 1}\sum_{i=0}^{N_c - 1}  \log_2(1+\gamma_{il}^u ) , m = 1, \cdots, M .
\end{equation}

The packet-level abstraction, where time is discretized into TTIs, is adopted. Each packet is scheduled within a TTI, and its successful transmission is determined by its communication rate. In particular, a packet transmission for a user is considered successful if the communication rate exceeds a predefined threshold $R_{\mathrm{th}}$. \textcolor{black}{Hence, the success transmission affected by the beam and power allocation is denoted by}
\begin{equation}
      \color{black} \mu_u[n] \triangleq \mu_u (\vect{V}[n],\vect{P}[n]) = \mathbbm{1}_{ \sum_{m=0}^{M-1} \mathrm{R}_{u,m} \ge R_{\text{th}}} .
\label{eq:communication_criterion}
\end{equation}
As such, the transmission process is modeled as a binary outcome, either success or failure, within each TTI, instead of a variable transmission duration. This abstraction enables us to focus on the impact of beam allocation on communication performance as well as queue evolution.

Backlog of untransmitted packets, as well as newly arrived packets, are accumulated in BS buffers for each user. Let $q_u[n]$ be the queue length, i.e., buffer occupancy, at the beginning of TTI $n$. \textcolor{black}{The queue length of the following TTI that captures the finite buffer size is given by
\begin{equation}
     q_u[n+1] = \min (\tilde{q}_u[n+1], B_u) ,
\end{equation}
where
\begin{equation}
    \tilde{q}_u[n+1] = \max(q_u[n] + a_u[n] - \mu_u[n] - \varsigma_u[n], 0) ,
\end{equation}
$a_u[n] \sim \text{Bernoulli}(P_u^{ar})$ is the arriving packet following Bernoulli distribution, $\varsigma_u[n] = \mathbbm{1}_{\mu_u[n]=0 \land \delta_u[n]= D_u}$ denotes the drop of failed transmitted packet whose delay exceeds the deadline.}

The communication module of this ISAC system aims to achieve low-latency transmissions. In other words, the BS attempts to minimize the average delay of all the users. When the buffers are infinite and packets have no deadline constraints, according to Little's law, the average queuing delay of user $u$ is proportional to the average queue length, written as $\lim_{N \rightarrow \infty} 1/N\sum_{n=1}^{N} q_u[n] / P_u^{ar}$ \cite{xie_noma_isac_allocation_2025}. 
However, while taking the buffer size as well as transmission deadline into account, it is challenging to obtain a closed-form expression of the average delay. 

To quantify the latency in the finite-buffer system with packet deadline constraints, we propose an online surrogate for the average delay. It is evaluated incrementally through direct system observation, avoiding a closed-form expression. As defined in Section \ref{sec:traffic_model}, $\delta_u[n]$ denotes the head-of-line packet latency in the $u$-th buffer at $n$-th time instant. 
\textcolor{black}{Its evolution can be given by
\begin{equation}
    \delta_u[n+1] =
    \begin{cases}
        \delta_u[n]+1, & \mu_u[n] = 0 \land \delta_u[n] < D_u\\[4pt]
        0, & \mu_u[n] = 1 \land q_u[n+1] = 0\\[4pt]
        \shortstack[l]{delay of the\\[-3pt] next packet,} &
        \shortstack[l]{$(\mu_u[n] = 1 \land q_u[n+1] \neq 0)\ \lor$\\[-2pt]
        $(\mu_u[n] = 0 \land \delta_u[n] = D_u)$}
    \end{cases}
\end{equation}
Thus, $\delta_u[n]\!\cdot\!\mu_u[n]$ provides a surrogate for the instantaneous queuing delay of a successfully transmitted packet. In addition, to accommodate the constraints of finite buffer size and packet deadline, we add an extra penalty to the delay. Then, the communication metric is the sum of the average empirical delays of successfully transmitted packets and penalties, written as%summing the average of $\delta_u[n]$ over $N$ observations and the penalty. 
\begin{equation}
    \bar{\delta}_u = \frac{1}{N}\sum_{n=1}^{N} \left(\delta_u[n]\!\cdot\!\mu_u[n] +  D_u \!\cdot\! \mathbbm{1}_{\delta_u[n] = D_u} + \mathbbm{1}_{\tilde{q}_u[n+1]>B_u} \right).
\label{eq:average_delay}
\end{equation}
}

\subsection{Sensing Model and Metric}
The transmitted signal impinges on the users and is reflected back to the BS sensing receiver. The BS utilizes this echo signal to estimate the range and velocity of each user, which can then be used to monitor traffic and/or further enhance communication. In this subsection, we outline the estimation process and then give the sensing metric.

The echo signal received on the $i$-th subcarrier and $l$-th OFDM symbol is written as
\begin{equation}
\begin{aligned}
     \vect{y}_i[l] =&  \vect{H}_{i,l} \vect{x}_i[l] + \vect{w}_i[l] \\
    =& \sum_{u=1}^{U} \beta_u \vect{a}^*(\phi_u)  \vect{a}^H(\phi_u)  e^{j4\pi i\Delta f \frac{d_u}{c} } e^{j 4 \pi l T_s v_u \frac{f_c}{c}} \vect{x}_i[l] \\ & \qquad + \vect{C}_i[l]\vect{x}_i[l] + \vect{w}_i[l] 
    %=& \sum_{u=1}^U \beta_u \vect{a}(\phi_u)  \vect{a}^H(\phi_u)  e^{ji \varphi_{u} } e^{j l \psi_{u}} \vect{x}_i[l] + \vect{w}_i[l]
\end{aligned}
\label{eq:echo_FD}
\end{equation}
where $\vect{H}_{i,l}$ denotes the total channel between the transmitter and the sensing receiver, $\vect{w}_i[l]\sim\mathcal{CN}(0,\sigma_\omega^2\vect{I})$ is the AWGN at the sensing receiver, $\beta_u$ is the complex reflection coefficient containing both the round-trip path-loss and the radar cross section of the user $u$. Similarly, $\vect{C}_i[l] = \sum_{c=1}^{C} \beta_c \vect{a}^*(\phi_c) \vect{a}^H(\phi_c) e^{j 4 \pi i \Delta f \frac{d_c}{c}}$ denotes channel reflected from $C$ stationary clutters, where $\phi_c$, $d_c$ and $\beta_c$ are AoD, distance to the BS, and the reflection coefficient of the $c$-th clutter. It is noted that $\vect{C}_i[l]$ is a constant since clutters are stationary.
We denote $\varphi_u = 4 \pi \Delta f d_u/c$ and $\psi_u = 4 \pi T_s v_u f_c/c$ the phase shifts induced by round-trip delay and Doppler, respectively, and they can be regarded as normalized angles that are analogous to the normalized AoD $\phi_u$. 

Various approaches have been proposed in the literature for estimating angle, range, and velocity information from the sensing echo. Typically, angles can be first extracted using conventional methods, such as MUltiple SIgnal Classification (MUSIC) and DFT, followed by a two-dimensional (2D) DFT to obtain range and velocity estimates. A three-dimensional (3D) extraction method was recently proposed in \cite{xiao_novel_2024}, which eliminates the impact of random signals by leveraging the known transmitted signal and finds estimates in formulated 3D cubes. This method focuses on the scenario with separated communication and sensing modules. In this case, the BS does not require user-beam association. To accommodate the shared communication and sensing applications that require user-beam association, we apply the signal processing approach in \cite{xiao_novel_2024} with slight modifications for the sensing module.

We propose reusing the transmitted beams for receiving echoes. Since the DFT codebook is utilized for transmit beamforming, it is equivalent to applying the DFT method and extracting the reflected echo in associated angular bins. Let us consider an example where the user $u$ is assigned with one beam and its AoD $\phi_u$ is within the mainlobe of the allocated beam. Denote the allocated beam as $[\Tilde{\vect{F}}]_{n_a}$, where $n_a$ is the index of the angular bin.
Then, the component extracted from the angular bin is expressed as 
\begin{subequations}
\begin{align}
    & {\mathcal{Y}}_{i,l}'[n_a] = [\Tilde{\vect{F}}]_{n_a}^T \vect{y}_i[l] \\ 
    =&  s_i^u[l] \beta_u |\vect{a}^H(\phi_u) [\Tilde{\vect{F}}]_{n_a}|^2  e^{j i \varphi_u }  e^{j l \psi_u}    \label{eq:angular-bin-user} \\
    & + \sum_{\forall u', \phi_{u'} \notin n_a} \beta_{u'} [\Tilde{\vect{F}}]^T_{n_a} \vect{a}^*(\phi_{u'}) \vect{a}^H(\phi_{u'}) e^{j i \varphi_{u'} } e^{j l \psi_{u'}} \vect{x}_i[l] \nonumber \\ &\quad\quad + [\Tilde{\vect{F}}]^T_{n_a} \vect{C}_i[l] \vect{x}_i[l]+ [\Tilde{\vect{F}}]^T_{n_a} \vect{w}_i[l] \nonumber  .
\end{align}
\end{subequations}
The first term in \eqref{eq:angular-bin-user} is the extracted echo of the target. The second term in \eqref{eq:angular-bin-user} comprises the contributions of other users and clutters in the reflected echoes, which, for simplicity, is denoted as $C_{i,l}[n_a]$ hereafter.

To estimate the range and the velocity from the extracted echo in an angular bin, the fast-varying $s_i^u[l]$ affects the following estimation. Owing to the co-located sensing receiver, this term can be removed by using its available full information, that is \cite{xiao_novel_2024},
\begin{equation}
\begin{aligned}
        \mathcal{Y}_{i,l}[n_a] =  \begin{cases}
            &{\mathcal{Y}}_{i,l}'[n_a] /s_i^u[l] ,s_i^u[l] \neq 0 \\
            &{\mathcal{Y}}_{i,l}'[n_a] , s_i^u[l] = 0
        \end{cases} 
\end{aligned}
\end{equation}
% Moreover, the extracted echo power averaged over subcarriers and OFDM symbols can be written as $\mathrm{E}\{|{\mathcal{Y}}_{i,l}[n_a]|^2\}$.
% while $I$ denotes the clutter interference, and we define the beamforming gain as 
% \begin{equation}
%     |\vect{a}^H(\phi) \Tilde{\vect{F}}| = 
%     \begin{cases}
%         G_M ,& \phi \in \mathcal{I}_M \\
%         G_S,& \phi \in \mathcal{I}_S
%     \end{cases}
% \label{eq:gain_lobes}
% \end{equation}
% where $\mathcal{I}_M$ and $\mathcal{I}_S$ are the mainlobe and the sidelobe of beam $\Tilde{\vect{F}}$ respectively, and $G_S \ll G_M$.

Given $\mathcal{Y}_{i,l}[n_a]$, the range and velocity can be estimated by searching for the peak in the range-Doppler bins. This is done by calculating the 2D-DFT of non-zero elements of $N_cN_s$ samples of $\mathcal{Y}_{i,l}[n_a]$ to transform it to the range-Doppler domain. \textcolor{black}{Let $n_r$ and $n_v$ denote the index of range bin and Doppler bin, respectively, where $n_r=1, \cdots, N_c$, and $n_v = 1, \cdots, N_s$. 
The normalized 2D-DFT of the $n_r, n_v$-th bin is given by }\cite[Chap. 4]{signal_and_system}
\begin{equation}
    \frac{1}{N_sN_c}\sum_{i=0}^{N_c -1} \sum_{l=0}^{N_s -1} \mathcal{Y}_{i,l}[n_a] e^{-j i \frac{2\pi n_r}{N_c}} e^{-j l \frac{2\pi n_v}{N_s}} .
\end{equation}
Therefore, to search for the peak of user $u$, the number of bins can vary from $N_cN_s$ to $3N_c N_s$ depending on the number of allocated transmit beams.
Denote the indices of the range bin and of the Doppler bin of the amplitude peak as $n'_r$ and $n'_v$, respectively. The range and velocity can be estimated as
\begin{equation}
    \hat{d} = \frac{c n'_r}{2 N_c \Delta f}, \quad \hat{v} = \frac{c n'_v}{2 N_s T f_c} .
\end{equation}

The achievable resolution of range and velocity estimation is dependent upon the number of respective bins, subcarrier spacing, and symbol duration, which are defined as $$\Delta_d = \frac{c}{2N_c \Delta f}, ~~ \Delta_v = \frac{c}{2f_c N_s T_s}.$$

The SINR of the extracted echo of target $u$ on a subcarrier of one symbol, expressed as
\begin{equation}
    \gamma_{n_a} \triangleq \gamma_{n_a}(\vect{V}[n],\vect{P}[n]) = \frac{P_u\left|\beta_u \right| ^2 \left|[\Tilde{\vect{F}}]_{n_a}^H \vect{a}(\phi_u) \right|^4 }{\left|C_{i,l}[n_a] \right|^2 + \sigma_n^2} ,
\label{eq:echo_sinr}
\end{equation}
\textcolor{black}{is impacted by the power and beam allocations. It determines the signal and the interference level in the range-Doppler bins, and thus, has a direct impact on the performance of  estimating range and velocity.}
Given the SINR, the Cramér-Rao lower bound (CRLB) of range, velocity, and AoD in the case of a single target are given by \cite{keskin_limited_2021}
\begin{equation}
\begin{aligned}
    &\mathrm{E} \{(\hat{d}_u - d_u)^2\} \geq \frac{3c^2}{8 N_sN_c\gamma_{n_a} \pi^2  \Delta f^2 (N_c^2-1)} \\
    &\mathrm{E} \{(\hat{v}_u - v_u)^2\} \geq \frac{3 \lambda^2}{8 N_sN_c\gamma_{n_a} \pi^2 T_s^2 (N_s^2-1)} \\
    & \mathrm{E}\{(\hat{\phi}_u - \phi_u)^2\} \geq \frac{6 }{N_sN_c\gamma_{n_a} \pi^2 \cos^2 (\phi_u) (N_R^2 - 1)} \triangleq \sigma^2_{\phi_u} %\label{eq:crlb_aod}
\end{aligned}
\label{eq:crlb}
\end{equation}

It can be observed that the estimation performance bounds presented in \eqref{eq:crlb} are functions of the SINR of the extracted signal using transmitted beams, i.e., $\gamma_{n_a}$.
Available works on resource management typically take CRLB or SINR as the sensing metric. These two metrics are functions of the transmit power for user $u$, and the gain achieved by the allocated transmitting beamforming vector $|[\Tilde{\vect{F}}]_{n_a}^H \vect{a}(\phi_u) |^4$. To maximize the beamforming gain when the transmitted power is fixed, multi-beams, where the power is spread, will be avoided. On the other hand, in practice, multi-beams can be useful to provide a wide FoV. As indicated by \eqref{eq:crlb}, the beamforming gain achieved by the allocated transmitting beams decreases when the target user is not in the center of the beam. Due to the beam coverage, the precision decreases. A wide sensing FoV helps alleviate the problem. For instance, the authors in \cite{zhang_multibeam_2019} proposed a multi-beam ISAC scheme that generates one time-invariant communication beam and one fast-varying sensing beam sweeping the area of interest during a channel coherence time. However, this requires the transmitter to rapidly switch the beamforming coefficients, thereby decreasing energy efficiency. Therefore, we propose to use multiple beams simultaneously for sensing to broaden the FoV. With limited beam resources, the number of beams assigned to each user must be optimized to achieve satisfactory performance for both the communication and sensing modules. In this work, instead of the general sensing metrics, we use the specific estimation errors normalized with respect to their resolutions. Specifically, for user $u$, the normalized estimation errors are given by 
\begin{equation}
    \begin{aligned}
        \epsilon^v_u[n] = \frac{|\hat{v}_u[n] - {v}_u[n]|}{\Delta_v} , ~~
  \epsilon^d_u[n] = \frac{|\hat{d}_u[n] - {d}_u[n]|}{\Delta_d} .
    \end{aligned}
    \label{eq:estimation_error}
\end{equation}
\textcolor{black}{The respective estimation errors averaged over $N$ TTIs are
\begin{equation}
    \bar{\epsilon}^v_u = \frac{1}{N}\sum_{n=1}^{N}\epsilon^v_u[n], ~~ \bar{\epsilon}^d_u = \frac{1}{N}\sum_{n=1}^{N} \epsilon^d_u[n]
\label{eq:average_estimation_errors}
\end{equation}}
% \textcolor{black}{The realized estimation error is influenced by the accuracy of peak search in the range-Doppler bins of which the noise level are impacted by the SINR of the extracted echo in \eqref{eq:echo_sinr}, which are, in turn, determined by the power and beam allocation.} 

\section{Beam Management in ISAC Systems} \label{sec:objective}
In this paper, we investigate the cross-layer beam management in ISAC systems. The BS allocates beams for multiple users to jointly satisfy the communication and sensing criteria. The communication metric is to minimize the average communication latency, while the sensing metric is to minimize estimation errors in range and velocity. The system objective function can be written as
% \begin{equation}
% \begin{aligned}    
%     \underset{\vect{V}, \vect{P}}{\text{minimize}} &~~ \sum_{u=1}^{U} \bar{\delta}_u + \epsilon^d_u + \epsilon^v_u \\
%     \text{subject to}&~~  \vect{P} = \diag(P_1, \cdots, P_U),  \\
%     &~~ N_s N_c M \trace(\vect{P}) \le P_{tot} \\
%     &~~ \|\vect{v}_u\|_0 \le 3, ~u = 1, \cdots, U \\
%     &~~ \vect{V}^T\vect{V} \text{~is diagonal} 
% \end{aligned}    
% \end{equation}
\textcolor{black}{
\begin{equation}
\begin{aligned} 
    \underset{\{\vect{V}[n], \vect{P}[n]\}_{n=1}^{N}}{\text{minimize}} &~~ \sum_{u=1}^{U} \bar{\delta}_u + \bar{\epsilon}^d_u + \bar{\epsilon}^v_u \\
    \text{subject to}~~&~~ \eqref{eq:average_delay}, \eqref{eq:average_estimation_errors}\\
    &~~  \vect{P}[n] = \diag(P_1[n], \cdots, P_U[n]),  \\
    &~~ N_s N_c M \trace(\vect{P}[n]) \le P_{tot}, \forall n \\
    &~~ \|\vect{v}_u[n]\|_0 \le 3, ~u = 1, \cdots, U , \forall n\\
    &~~ \vect{V}^T[n]\vect{V}[n] \text{~is diagonal}, \forall n 
\end{aligned}    
\end{equation}}
\begin{remark}
% Even though the optimization variables do not directly appear in each term of the objective function, they exert implicit impacts through system dynamics. In particular, beam allocation and power decisions jointly determine the communication rate and estimation sensing quality, which in turn affects the queue evolution and estimation accuracy. The implicit coupling between the objective function and the optimization variables renders the optimization problem analytically intractable and challenging to solve by conventional approaches. 
\textcolor{black}{The components of communication quality and sensing quality in the objective functions are given in \eqref{eq:average_delay} and \eqref{eq:average_estimation_errors}, respectively. Closed-form expressions of these objectives as functions of beam and power allocation over $N$ TTIs, i.e., $\{\vect{V}[n], \vect{P}[n]\}_{n=1}^{N}$, cannot be found. However, they are directly depend on them. Specifically, $\vect{V}[n]$ and $\vect{P}[n]$ determine the communication SINR in \eqref{eq:sinr_user_signal}, the communication rate, and hence the indicator of success transmission in \eqref{eq:communication_criterion}. Then, the queue and head-of-line delay evolution set the empirical delay $\bar{\delta}_u$. 
With respect to the sensing objectives, the same beam and power allocation decide the extracted echo SINR in \eqref{eq:echo_sinr}, which affects the range-Doppler peak search and consequently, the accuracy of the range and velocity estimation. The implicit coupling between the objective function and the optimization variables renders the optimization problem analytically intractable and challenging to solve by conventional approaches. 
}
\end{remark}

To reduce latency and estimation error, the number of beams and their directions need to be determined and refined in time. This typically requires the BS to have downlink channel information via user feedback or estimation during pilot transmission. To reduce the overhead associated with channel information acquisition, the BS resorts to the sensing module in the ISAC schemes.

\subsection{AoD-Based Beam Management}
\label{sec:aod-base}
We first propose a heuristic method in which the BS dynamically determines the number of beams for each user at each TTI. Note that the buffer state is not taken into account.

The BS estimates the AoDs of all users using the MUSIC algorithm. Then, it determines the number of beams per user based on the AoD estimation precision, since the precision implies the achieved SINR. That is, when the estimation error of AoD for a user is large, the BS needs to allocate multiple beams to increase the FoV and improve the likelihood of finding the user's AoD. 
Therefore, the BS makes the beam allocation decision by comparing the estimation precision, $\sigma_{{\phi}_u}$, to a predefined threshold, $\epsilon_a^*$.  
Specifically, when $\sigma_{{\phi}_{u}} \ge \epsilon_a^*$, the BS allocates multiple beams to the user, namely the beam corresponding to the estimated AoD together with two adjacent beams. On the other hand, when $\sigma_{{\phi}_{u}} < \epsilon_a^*$, the estimation error is within an acceptable range, and the BS keeps the same beam allocation to the user. 
In the implementation, the estimation precision $\sigma_{{\phi}_u}$ is the standard deviation of the AoD estimate, computed from the CRLB in \eqref{eq:crlb}. 
To circumvent this, the AoD-based method uses an approximated CRLB by substituting AoD estimates.

\subsection{Power Allocation}\label{sec:power_allocation}
Once the beams for a given TTI are allocated, the BS computes the transmit power for the users on each subcarrier. Since all users are assumed to use the same modulation and coding scheme, the communication requirements can be expressed as per-subcarrier SINR targets for each user. Specifically, the effective channel and interference channel are obtained from the angle and range estimates of the previous TTI. 
When the SINR targets are feasible, the scheme with minimum total required power that meets the targets is chosen. If no feasible power vector exists within the power budgets, the BS resorts to a simple fallback strategy in which the available power is evenly divided among users on that subcarrier. Users that still fail to meet their SINR target under this fallback are marked as transmission failure and handled accordingly by the beam scheduling in the next TTI.

\begin{figure}
    \centering
    \includegraphics[width=0.93\linewidth]{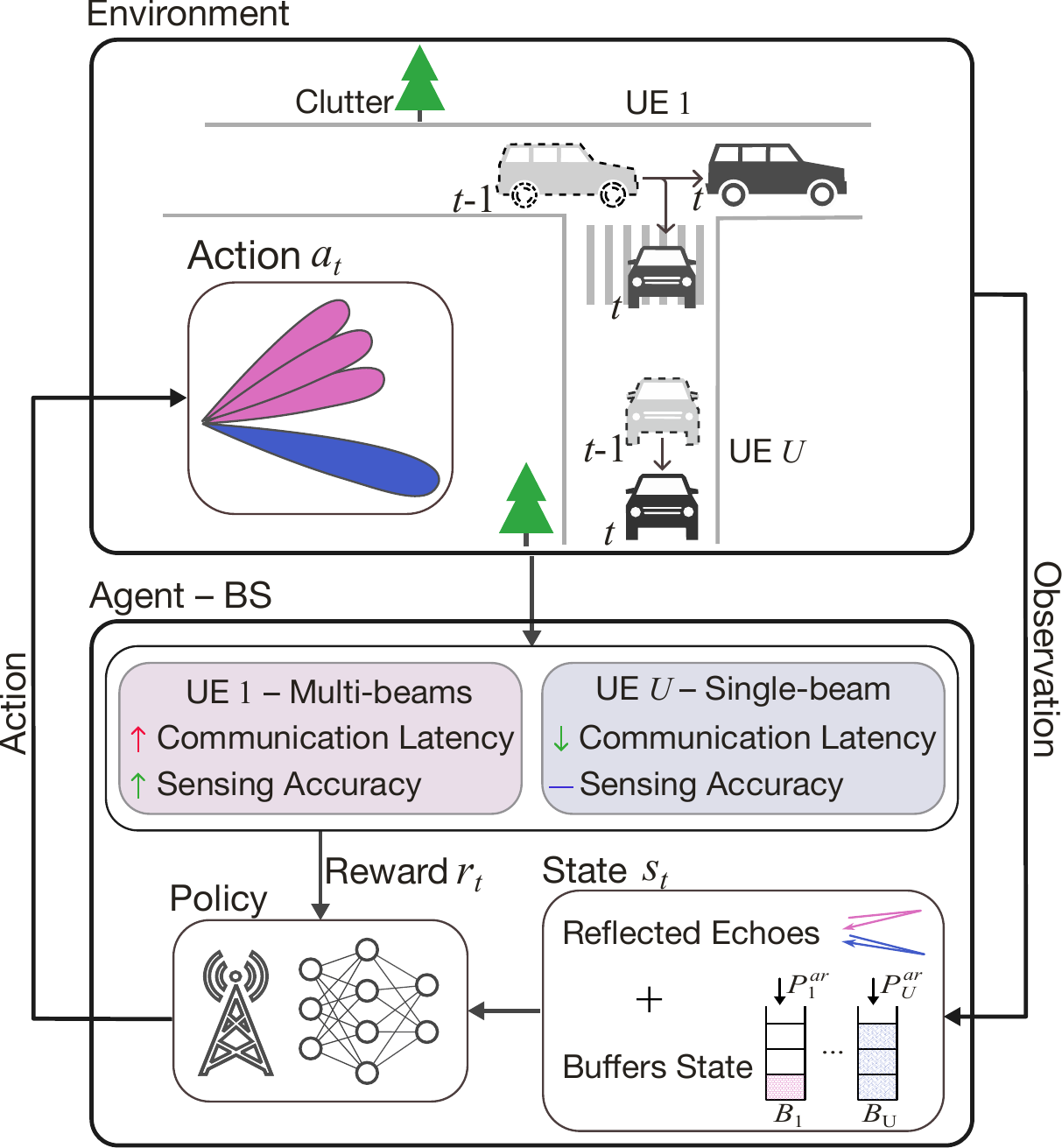}
    \caption{An illustration of DRL-assisted cross-layer beam allocation for ISAC systems. }
    \label{fig:drl_bm}
\end{figure}

\section{DRL-Assisted Cross-Layer Beam Management}
\label{sec:drl_method}
% While existing approaches to cross-layer resource management in ISAC systems have shown promise, they often rely on channel state information to achieve optimal performance. 
%For instance, \cite{xie_noma_isac_allocation_2025} proposes a stochastic scheduling scheme which derive the closed-form solution for rate adaption given quantized channel state information. Acquiring this information, however, introduces substantial control signaling overhead that can degrade system efficiency. Moreover, integrating buffer state information into analytical problem formulations presents additional complexity, as the inter-dependencies between queue dynamics, channel conditions and beam allocation decisions are difficult to capture in tractable closed-form expressions. In practical ISAC deployments, obtaining accurate and timely channel state information becomes increasingly challenging due to the dynamic nature of wireless environments and the dual sensing-communication functionality. 
We propose a DRL approach that performs intelligent beam allocation without requiring explicit channel knowledge. A schematic is illustrated in Fig. \ref{fig:drl_bm}. The DRL agent (BS) leverages only the reflected echoes at the sensing receiver as well as the buffer states to make adaptive beamforming decisions. The DRL approach eliminates the control signaling burden while naturally accommodating complex cross-layer interactions that are difficult to model analytically, making it particularly suitable for dynamic ISAC scenarios where traditional model-based methods may be impractical or suboptimal. 

The beam allocation problem of cross-layer ISAC systems is formulated as a Markov decision process (MDP). An MDP consists of a state space $\mathcal{S}$, an action space $\mathcal{A}$, an initial state distribution $p(s_1)$, a state transition distribution satisfying the Markov property $p(s_{t+1}| s_t, a_t) = p(s_{t+1}|s_t, a_t, \cdots, s_1, a_1)$, and a reward function $r: \mathcal{S} \times \mathcal{A} \rightarrow \mathbb{R}$.

\subsection{DRL Architecture for ISAC Beam Management}
\label{subsec:ppo_architecture}
In order to train a DRL agent, the observations, actions, and reward function are identified in this subsection. 
\subsubsection{Observations}
As discussed in Section \ref{sec:objective}, the beamforming gain achieved by the transmitted beamforming vector implicitly indicates the beamforming quality. To obtain an overview of the beam allocation quality and users' moving directions, the BS calculates the beamformed echo power using all the beamforming vectors in the codebook. 
The input to the learning agent is a vector of dimension $2N_T + 2U$ with real-valued entries, containing the reflected signal power in $N_T$ angular bins, the allocation from the previous step, and the buffer states of all users. The buffer states of $U$ users are the number of packets in their respective buffers and the longest waiting time. At each time step $t$,  the state described in the MDP is expressed as
\begin{equation}
    s_t = \{\vect{b}_1, \cdots, \vect{b}_{N_T}, \bar{\vect{v}}^{(t)}, \delta_1, \cdots, \delta_U, B_1, \cdots, B_U\} \in\mathcal{S}
\label{eq:state}
\end{equation}
where $\vect{b}_{n_a} = \mathrm{E}\{|\mathcal{Y}'[n_a]|^2\}, n_a = 1, \cdots, N_T$, and the expectation is taken with respect to subcarriers and OFDM symbols of a TTI,  $\bar{\vect{v}}^{(t)}$ is a $N_T \times 1$ vector with each entry indicating the specific beam is idle, equals to 0, or is allocated to user $u$, $\delta_u$ is the longest waiting time of a user's buffer and $B_u$ denotes the buffer occupancy.
\subsubsection{Actions}
Ideally, the actions taken by the agent are $U$ discrete values referring to the beam indices allocated to each user. It is not straightforward to determine the dimension of the action space $\mathcal{A}$ because each user may have one or multiple beams. One alternative is to let the agent output a vector of dimension $N_T$, with each entry indicating which user it is allocated to or staying idle. This results in an action space dimension of $(U+1)^{N_T}$. In the mmWave spectrum, the number of beams $N_T$ is large, which can hinder convergence of the learning algorithm and requires substantial memory. Thus, we shrink the action space by defining the action as follows. 
For the user $u$, let $\mathcal{C}_{u,t}$ denote the beam indices that are allocated for user $u$ at step $t$. The beam allocation is 
\begin{itemize}
    \item $a_u = 1$, beams stays the same $\mathcal{C}_{u, n} = \mathcal{C}_{u, n-1}$
   \item $a_u = 2$, the agent decides the best one based on the previous allocation
   $\mathcal{C}_{u, n} = \arg \max \vect{b}_{\mathcal{C}_{u, n-1}}$.
   \item $a_u = 3$, beams sweep to the area with smaller indices, $\mathcal{C}_{u, n} = \min(\mathcal{C}_{u, n-1}) + [-2,-1,0]$
   \item $a_u = 4$, the beam switches to the adjacent smaller one, $\mathcal{C}_{u, n} = \min(\mathcal{C}_{u, n-1}) - 1$ 
 \item $a_u = 5$, beam sweep to the area with bigger indices, $\mathcal{C}_{u, n} = \max(\mathcal{C}_{u, n-1}) + [0, 1, 2]$
 \item $a_u = 6$, the beam switches to the adjacent bigger one, $\mathcal{C}_{u, n} = \max(\mathcal{C}_{u, n-1}) + 1$ 
\end{itemize}
As a result, the action space dimension decreases to $6^U$.
\subsubsection{Reward}
The reward at each time instant $t$, $r_t$, is composed of the communication performance and the sensing performance
\begin{equation}
    r_t =\sum_{u = 1}^{U}  (D_u - \delta_u) - \mathbbm{1}_{B_u = B_u^*} - \mathbbm{1}_{\delta_u = D_u} + \log_{b}\epsilon^d_u + \log_{b} \epsilon^v_u .
\label{eq:reward}
\end{equation}
The first term evaluates the latency of a packet. The second and the third terms serve as penalization factors for drops induced by buffer overflow and packet expiration, respectively.  The last two terms evaluate the estimation errors of range and speed. The reason for choosing the logarithm is to prevent reward explosion when $\epsilon$ is too large, and it gives a positive reward when the normalized estimation error is below a resolution threshold. The logarithm base $b$ indicates the weight of the sensing module in the reward function.

In the multi-user scenario, to reduce inter-beam interference, beams cannot be reused.
In case of beam allocation collision, several scheduling schemes have been considered, such as round-robin, proportional fair, and SNR-based scheduling \cite{mabrouki_codebook_2022}. In this work, as latency is regarded as an important metric directly available at the BS, a buffer-prioritized strategy is adopted. The BS prioritizes the user with the fuller buffer in the event of a collision. The user with lower priority gets the adjacent beams. 

\subsection{RL Algorithm Selection}

With the MDP formulation established, we now discuss the selection of the RL algorithms. Although model-based RL methods are efficient, they require the agent to learn an accurate world model of the environment dynamics. Regarding to cross-layer ISAC beam management, a model-based approach would entail modeling the joint evolution of user mobility, the angle-of-arrival estimation process, and the sensing-communication trade-off. These involve coupled physical dynamics that are challenging to learn from data alone. Hence, applying model-based RL to cross-layer beam management is a non-trivial extension that needs a dedicated investigation and is left for future work.

The beam management problem involves a discrete action space, which restricts the direct use of many DRL algorithms. For instance, among model-free DRL algorithms, soft actor-critic (SAC), deep deterministic policy gradient (DDPG), and the recently proposed TD7 \cite{td7_2023} rely on deterministic or stochastic policy gradients with respect to continuous action spaces. Applying these algorithms thus requires modifications by discretizing the continuous space in a piecewise manner. In addition, actor-critic (A2C), deep Q-network (DQN), and proximal policy optimization (PPO) are natively designed for discrete action spaces. DQN and TD7, with their off-policy mechanism, may learn from outdated beam allocation experiences. PPO addresses both concerns. It introduces a clipped surrogate objective to prevent excessively large policy updates, yielding more stable training. PPO's on-policy mechanism adapts to varying communication and sensing conditions at each update. These properties motivate us to mainly consider PPO for the cross-layer beam management. A numerical comparison between among PPO, A2C, DQN, and TD7 is provided in Section \ref{sec:simulation}.

\subsection{PPO Algorithm}
Proximal
policy optimization (PPO) \cite{ppo_2017} is a popular DRL algorithm that
prevents the policy from having a large update and performs
comparably or better than other DRL algorithms. As in other DRL algorithms, an agent is trained to optimize a policy $\vect{\pi}$ to maximize the return, denoted as $\mathbbm{r}_t(\gamma)$. A policy maps observed states to actions $\pi: \mathcal{S} \rightarrow \mathcal{A}$ and in turn obtains reward $r_t(s_t, a_t)$. The return $\mathbbm{r}_t(\gamma)$ is defined as the total discounted reward from the timestep $t$ and can be expressed as $\mathbbm{r}_t(\gamma) = \sum_{t' = t}^{\infty} \gamma^{t'-t} r(s_{t'}, a_{t'})$, where $\gamma \in[0,1]$, the discounting factor, captures the importance of the future rewards in the current value estimate. The objective of the agent is to find a policy $\vect{\pi}$ which maximizes the expected cumulative discounted return $J(\vect{\pi}) = \mathrm{E}[\mathbbm{r}_1(\gamma)|\vect{\pi}]$. The agent computes the value function of a policy $\vect{\pi}$ for each state as the expected return from that state, i.e., $V^{\pi}(s) = \mathrm{E}[\mathbbm{r}_1(\gamma)|S_1 = s; \vect{\pi}]$. The relative value of a selected action is estimated by the advantage function, defined as
$A^{\vect{\pi}}(s,a) =\sum_{i=t}^{N} \gamma^{i-t} r_i -V^{\pi}(s)$. It indicates how rewarding a particular action $a$ is taken from state $s$. 
At each timestep, the PPO agent takes an action that maximizes a clipped advantage function. The signature clip function 
\begin{equation}
\begin{aligned}
    L^{\text{CLIP}} (\theta) = \min & \left(  \frac{\vect{\pi}_{\theta}(a|s) }{\vect{\pi}_{\theta_t}(a|s)}A^{\pi_{\theta_t}}(s,a), \right.  \\ & \left. \text{clip}\Big(\frac{\vect{\pi}_{\theta}(a|s) }{\vect{\pi}_{\theta_t}(a|s)}, 1-\varepsilon, 1+\varepsilon \Big)A^{\pi_{\theta_t}}(s,a) \right) ,
\end{aligned}
\label{eq:clip_func}
\end{equation}
where $\theta$ is the neural network parameters, $\varepsilon$ is the clipping parameter, and the $\text{clip}(\cdot)$ function restricts the probability ratio to $[1-\varepsilon, 1+\varepsilon]$. Eq. \eqref{eq:clip_func} can be equivalently written as
\begin{equation}
\begin{aligned}
   L^{\text{CLIP}} (\theta) = 
   \begin{cases}
         \min \left(  \frac{\vect{\pi}_{\theta}(a|s) }{\vect{\pi}_{\theta_t}(a|s)}, ( 1+\varepsilon )  \right) A^{\pi_{\theta_t}}(s,a) , \\ \qquad \qquad  \qquad \quad \qquad A^{\pi_{\theta_t}}(s,a) \ge 0 , \\
         \max \left(  \frac{\vect{\pi}_{\theta}(a|s) }{\vect{\pi}_{\theta_t}(a|s)}, ( 1-\varepsilon )  \right) A^{\pi_{\theta_t}}(s,a), \\ \qquad \qquad \qquad \qquad \quad A^{\pi_{\theta_t}}(s,a) < 0 .
   \end{cases} 
\end{aligned}
\end{equation}
The function thus prevents excessively large policy updates.

% The policy can be either created by a tabular approach or by function approximation approach. In the case where the states and actions are discrete, tabular methods can be used to capture the value for $Q^{\vect{\pi}}(s, a)$, which in turn decides the action taken by policy $\vect{\pi}$. On the contrary, when states or actions take continuous values or their corresponding spaces have a high dimension, function approximation methods are utilized because keeping a table for Q-values becomes infeasible. By appropriately defining a family of functions that operate on $(s,a)$ and learning the parameters of the function which maximizes the return from the policy $\vect{\pi}$, these methods depend on the expressive power of candidate functions for the success of DRL policies. Multiple function approximation techniques including linear functions, radical basis functions, Fourier basis, neural networks, have been proposed. 

PPO, as a DRL algorithm, utilizes a deep neural network for the function approximation. The employed neural network can be regarded as a chain of functions that transforms its input into a set of outputs via a non-linear transform. In DRL, at each step, the agent uses the trained neural network policy to output an action $a$ after observing a state $s$. Using the clip function in \eqref{eq:clip_func}, PPO seeks to take the largest possible improvement step on a policy based on the current available data, without stepping too far, thereby avoiding a collapse in performance.

% The sensing performance metric can be set as estimation error of delay and Doppler using the 2D-DFT estimation approach. 

% In radar literature, it has been shown that mutual information (MI) can be used as a criterion. MI is defined between the received radar echoes and the target scattering coefficient. The MI criterion for sensing is widely used for waveform optimization in radar system. There is an information-theoretic relation between MI and MSE that could be used in tracking or parameter estimation \cite{guo_mutual_2005}. However, optimizing the MI for sensing can introduce an undesired peaks in the range of sidelobes.

% \section{Problem Formulation}
% The objective function of every DRL step is to minimize the packet loss and quantized MSE.
% \begin{equation}
%     \underset{\vect{V}}{\text{minimize}} \sum_{u=1}^{U} \mathbbm{1}_{\mathrm{SE}_u < R_{th}} + \left\lceil \frac{\mathrm{MSE}_{v_u}^{1/2}}{\Delta_v }\right\rceil + \left\lceil \frac{\mathrm{MSE}_{d_u}^{1/2}}{\Delta_d}\right\rceil
% \end{equation}

\begin{table}
    \centering
    \caption{SIMULATION PARAMETERS}
    \resizebox{0.4\textwidth}{!}{%
    \begin{tabular}{l l l}
    \hline \hline
    Description &Symbol &Value \\ 
    \hline
    BS Antennas &$N_T$& 16 \\
    Frequency &$f_c$& 28 GHz \\
    Subcarrier Spacing &$\Delta f$ &60 kHz \\
    FFT Size & $N_c$ & 144 \\
    Number of OFDM symbols &$N_s$ &280 \\
    OFDM Symbol Duration &$T$ &18.9 $\mu$s \\
    Tx Power &$P_t$& $7$ dBm \\
    Noise Power& $\sigma_{\omega}^2$& $-109$ dBm\\
    Rician K Factor& -- &10 dB\\
    Time Duration of a TTI& $dt$& 20 ms\\
    % Beamwidth - Sensing Threshold & $\theta_{bw}$ & $\approx 4^{\circ} $ \\
    %SE - Communication Threshold & $R_{\mathrm{th}}$ & 3 bits/Hz \\
   Radar Cross Section  & $\sigma_{rcs} $ & 100 $m^2$ (20 dBsm) \cite{smida_inband_2024}\\
    Scenario Dimensions & -- & $[100 \times 100]$ \\
    Number of UEs &$U$& 2 \\
    Packet Arrival Probability & $[P_1^{ar}, P_2^{ar}]$ & $[0.9, 0.7]$\\
    Buffer Size &$[B_1, B_2]$ & $[6, 8]$ \\
    Deadline & $[D_1, D_2]$ & $[6, 8]$ \\
    User Speed &$v_1, v_2$&$\sim \mathcal{N}(v,4), v=6,8,10,12,14$ \\
    Number of TTI per Frame &$N$ & 40 \\
    Speed estimation resolution & $\Delta_v$ & $1.01$m/s \\
    Range estimation resolution & $\Delta_d$ & $1.736$ m \\
    \hline \hline
    \end{tabular}%
   }
    \label{tab:exp_para}
\end{table}

\begin{figure}
    \centering
    \includegraphics[width=0.8\linewidth]{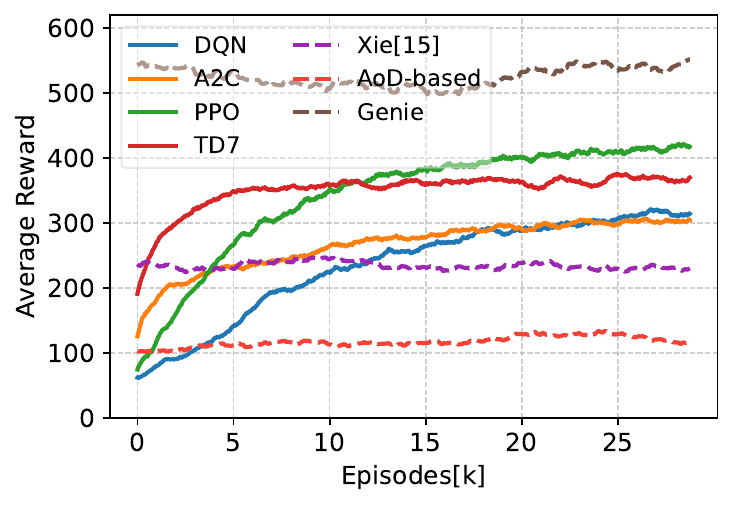}
    \caption{Comparison between the training reward of RL algorithms and the reward of benchmarks in the cluttered channel condition.}
    \label{fig:rl_comparison}
\end{figure}

\begin{figure*}
\centering
    \centering
    \setlength{\tabcolsep}{0pt}
    \begin{tabular}{cccc}
        \subfloat[]
    {\includegraphics[width=0.24\textwidth]{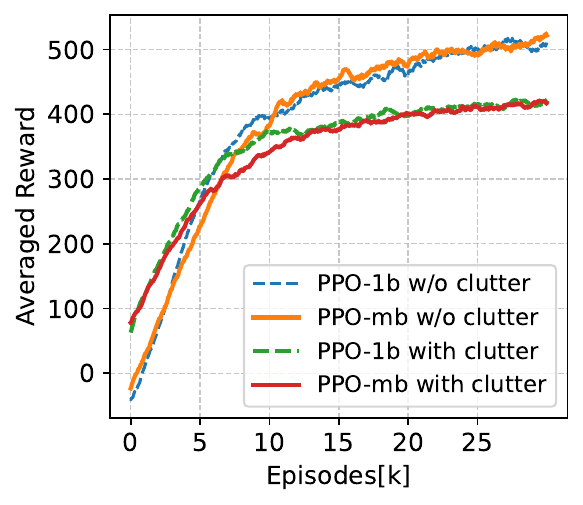}
    \label{fig:reward_1b_mb}}
    \subfloat[]
    {\includegraphics[width=0.24\textwidth]{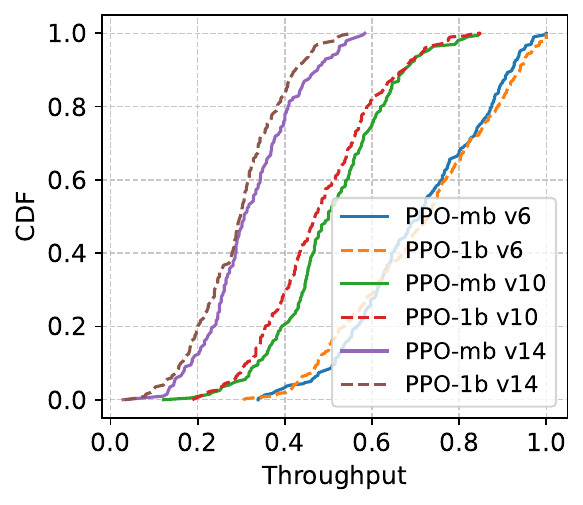}
    \label{fig:thp_1b_mb}} 
        \subfloat[]
    {\includegraphics[width=0.24\textwidth]{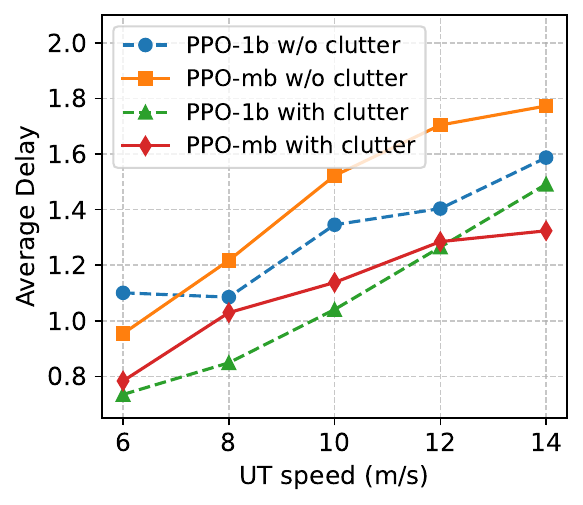}
    \label{fig:latency_1b_mb}} 
        \subfloat[]
    {\includegraphics[width=0.24\textwidth]{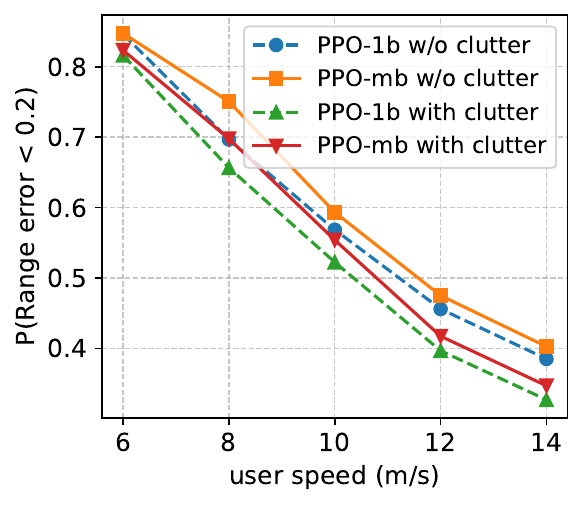}
    \label{fig:range_1b_mb}} 
    \end{tabular}
\caption{Performance of the proposed DRL-assisted method. The comparison between PPO-1b (one beam for a user) and PPO-mb (one or multiple beams for a user) in the clean and cluttered scenarios: In (a), training reward; (b), CDF of throughput in the cluttered condition when users velocities are $v=6, 10, 14$ m/s, (c) averaged latency and (d) range estimation error as a function of user speed.}
\end{figure*}

\section{Simulation Results and Discussion}
\label{sec:simulation}

This section presents the simulation results of the proposed multi-beam management for ISAC systems. The simulation parameters are summarized in Table~\ref{tab:exp_para}. Specifically, the considered area is a $100 \text{m} \times 100 \text{m}$ area with a BS situated at $[20,50]$. We consider two users; user 1 is closer to the BS than user 2. Two users are assumed to be vehicles with a radar cross section of 20 dBsm \cite{smida_inband_2024}. There are four intersections in the scenario. Upon reaching an intersection, a user turns in any of the three other directions with equal probability. Users' speed at a TTI follows a Gaussian distribution with mean $v = 6,8,10,12,14$ m/s and variance $4$.
The duration of a TTI is 20 ms. There are six sets of user initial positions. During testing, 500 episodes are evaluated, each with user positions initialized randomly from a test set that is disjoint from the training set.

We consider two channel conditions, with and without (w/o) clutter in the environment. When there is no clutter, the channels are assumed to be line-of-sight (LoS). When clutters are present, the channel between the BS and a user is assumed to have Rician fading with a Rician K-factor of 10. The channel seen at the BS sensing receiver includes the paths reflected from these clutters. Four clutter are set, with their polar coordinates with respect to the BS transmitter given by $[13, \pi/5], 
[20, 3\pi/14],
[29, 9\pi/17],
[19, 15\pi/17]$.  

We compare the proposed approach with some baselines described as follows:
\begin{itemize}
\item 
\emph{$X$-TDMA ($X$ Time Division Multiple Access)}: This scheme imitates how the standardized 3GPP beam alignment protocol operates. It is independent of the agent's environment. The BS assigns a multi-beam slot to all users to update the best beams. The slot is followed by $X$ single-beam slots. 

\item
\emph{Genie}: This sets the performance upper bound. This method assumes genie AoD information as well as their association with users. According to the genie AoD, the agent allocates the best beam to the specific user.

\item
\emph{AoD-based}: The heuristic AoD-based method proposed in Section \ref{sec:aod-base}. The BS estimates the AoDs of multiple users from the reflected echoes using the MUSIC algorithm. The method assumes a known association between the user and AoD. Based on the estimated AoDs, the BS allocates a single beam or multiple beams towards each user depending on the estimation error. 

\item
\emph{Xie\cite{xie_noma_isac_allocation_2025}}: Assuming available channel state information, the beamforming vector is optimized to satisfy the transmit power and the transmission rate for all users. 
\end{itemize}

We first provide a comparison between the DRL-assisted approach with different DRL algorithms including DQN, A2C, PPO and TD7 and the benchmarks. The action space of TD7 is uniformly quantized into $6$ discrete level, each representing an action index. The agents are trained for $1.2 \times 10^6$ steps, with each step equivalent to one TTI. The logarithm base $b$ in the reward function \eqref{eq:reward} is set to $1/3$. The strategy of step decay learning-rate scheduling is adopted to assist policy convergence. The moving average training rewards are depicted in Fig. \ref{fig:rl_comparison}. As observed, RL-assisted approaches exceed the non-RL benchmarks after convergence while Genie gives an upper bound. A2C and DQN yield similar rewards, while DQN suffers from a much slower convergence speed. TD7, which outperforms A2C and DQN, converges at approximately $7\text{k}$ episodes (each episode has 40 steps). Nevertheless, PPO surpasses A2C after more than $12\text{k}$ training episodes, despite its relatively slower convergence. Hence, we select PPO as the agent for the DRL-assisted approach hereafter.  

In order to evaluate the advantage of using multi-beams, we present a comparison between using a single beam and multi-beams for a user.
The proposed DRL-assisted beam management is termed as \emph{PPO-mb}. In addition, we implement \emph{PPO-1b},  in which the actions are $a_u = 1, 4, 6$ described in Section \ref{subsec:ppo_architecture}. In this case, the BS directs only one beam towards a user. The neural network shares the same architecture as PPO-mb. 
Fig. \ref{fig:reward_1b_mb} illustrates the training convergence of the two PPO architectures. The PPO-based methods with clutters start with a higher reward than those without clutters. This is because the reward is sensing-dominated during the initial training phase: the estimated range may correspond to clutters rather than users, but the estimation error is smaller than in the clutter-free case. It can also be observed that PPO-1b and PPO-mb achieve similar rewards at the end of the training. We then examine a more detailed performance comparison between the two. 

Fig. \ref{fig:thp_1b_mb} and Fig. \ref{fig:latency_1b_mb} show the communication performance by comparing the cumulative distribution function (CDF) of throughput and the average latency of transmitted packets, respectively. It can be observed from Fig. \ref{fig:thp_1b_mb} that PPO-mb outperforms PPO-1b as user speed increases. This shows that using multiple beams is a feasible solution for addressing beam misalignment. That said, multiple beams result in a longer average delay, as seen in Fig. \ref{fig:latency_1b_mb}, because the transmit power is spread across multiple beams, reducing spectral efficiency. However, the multi-beam scheme can update beams and transmit packets within the deadline constraints, thus achieving better throughput. Moreover, Fig. \ref{fig:range_1b_mb} shows the sensing performance using range estimation as an example. It depicts the probability that the range estimation error satisfies $\epsilon_u^d < 0.2$ as a function of user speed. The threshold is set according to the sensing key performance indicator defined in the 6G vehicular network use case \cite{hexa}. As can be seen, PPO-mb provides better sensing performance because multiple beams offer broader coverage. Therefore, using multiple beams provides better performance for ISAC systems, enabling them to accommodate more practical dynamic scenarios. 

\begin{figure*}
\centering
    \centering
	\setlength{\tabcolsep}{0pt}
	\begin{tabular}{ccc}
		\subfloat[]
    {\includegraphics[width=0.32\linewidth]{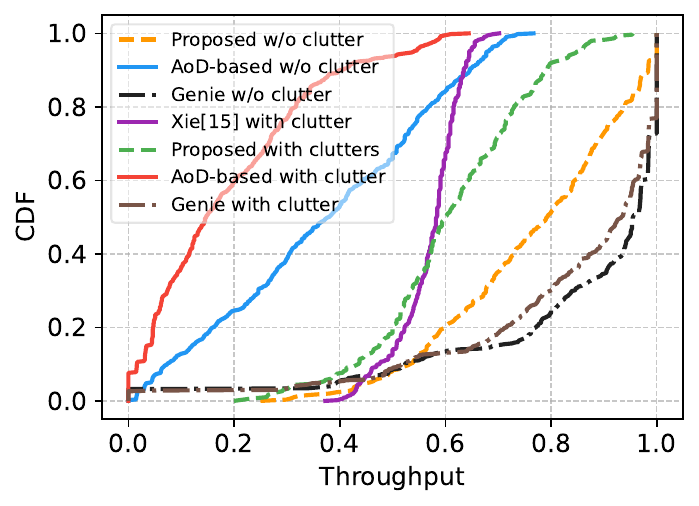}
    \label{fig:cdf_thp}} 
        \subfloat[]
    {\includegraphics[width=0.33\linewidth]{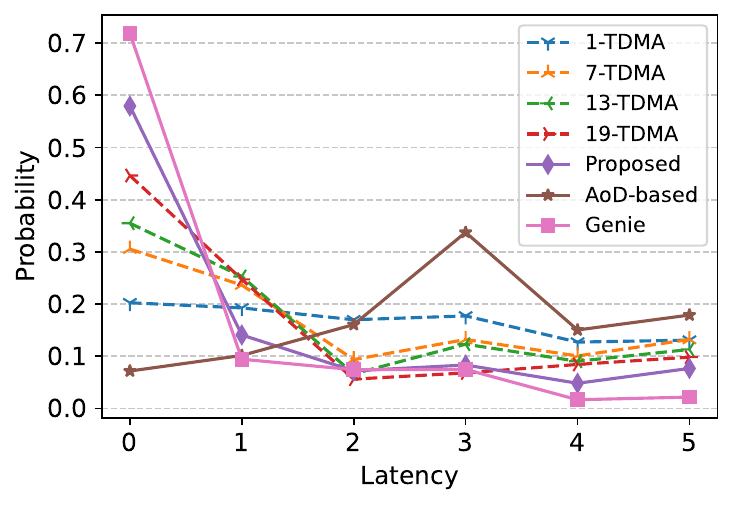}
    \label{fig:latency}}  
    \subfloat[]
    {\includegraphics[width=0.33\linewidth]{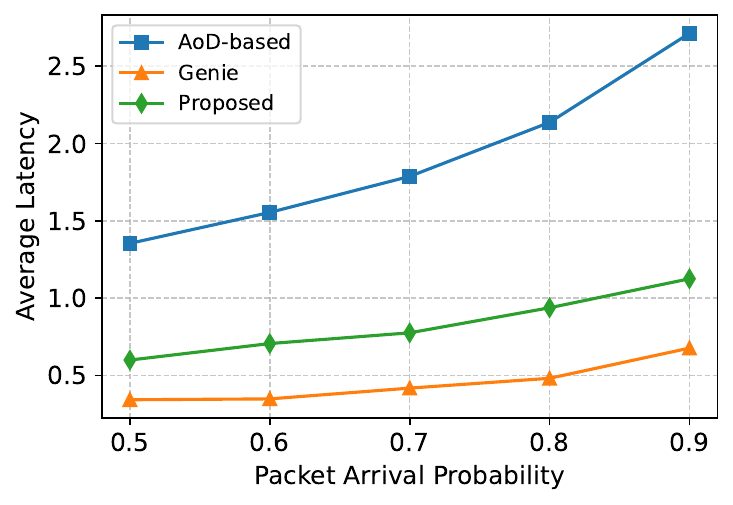}
    \label{fig:averaged_delay}}
    \end{tabular}
\caption{Communication performance results. In (a), comparison of the throughput CDF of different methods in the two channel conditions. In (b), a comparison of the latency of transmitted packets, i.e., the number of waiting TTIs before transmission, is made between different methods in the cluttered environment. In (c), Variation in the averaged latency of transmitted latency as a function of packet arrival probability.}
\label{fig:communication_performance}
\end{figure*}

In the following, we compare the communication and sensing performance of the various methods, considering only the \emph{PPO-mb} configuration for the proposed DRL-assisted approach. The average user speed is $v=6$ m/s.

\subsection{Communication Performance}
We first evaluate the communication performance. Fig.~\ref{fig:cdf_thp} presents the CDF of the throughput of all users compared among different methods. It is clear that the system without clutter achieves better throughput, as Rician fading in a cluttered environment reduces the SINR. Specifically, the AoD-based method suffers significant performance degradation because AoD estimation is strongly affected by clutters. The genie and the AoD-based method provide the upper and lower bounds, respectively. Using the proposed PPO-based method or Xie\cite{xie_noma_isac_allocation_2025}, the system can achieve 50\% of the throughput with more than 85\% confidence. The Xie~\cite{xie_noma_isac_allocation_2025} method yields stable throughput in the range of 0.37 to 0.7, but with an overall lower average. 

\begin{figure*}
\centering
	\centering
	\setlength{\tabcolsep}{0pt}
	\begin{tabular}{ccc}
		\subfloat[Range error (clutter)]{\includegraphics[width=0.33\textwidth]{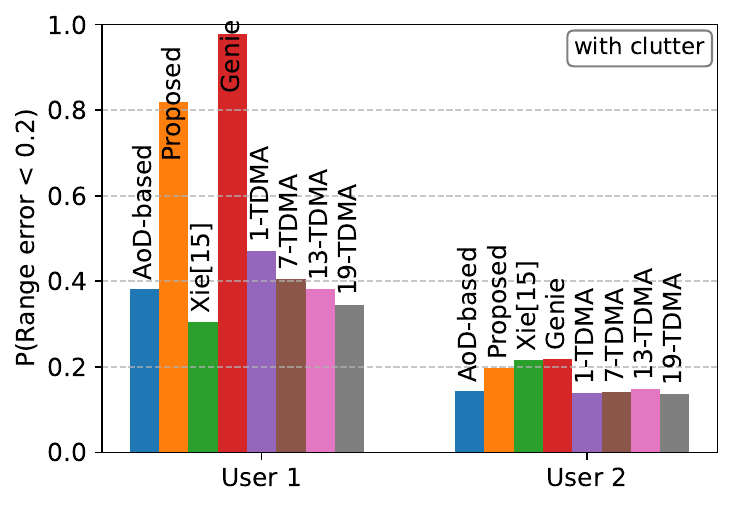} \label{fig:range_error_clutter}} 
		\subfloat[Velocity error (clutter)]{\includegraphics[width=0.33\textwidth]{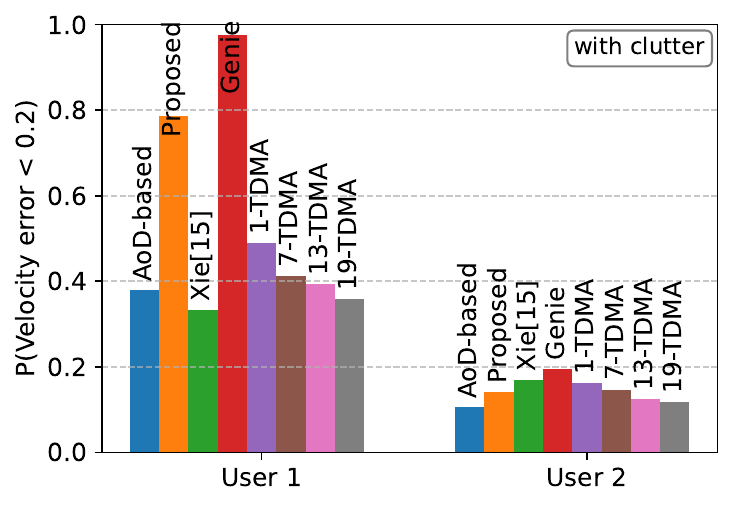}
	   \label{fig:velocity_error_clutter}} 
            \subfloat[Angle error (clutter)]{\includegraphics[width=0.33\textwidth]{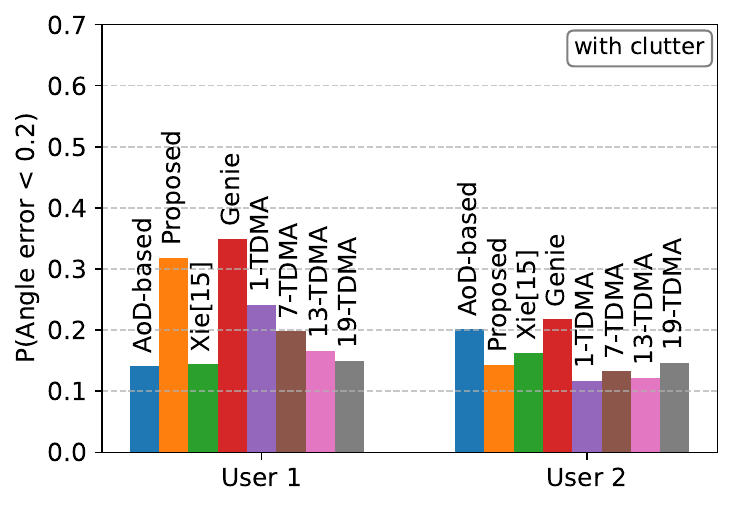}\label{fig:angle_error_clutter}}
       \\
            \subfloat[Range error (clear)]
        {\includegraphics[width=0.33\textwidth]{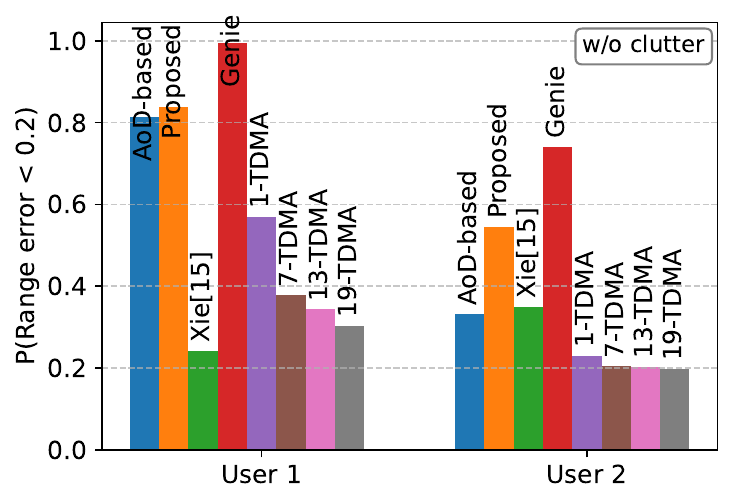}
			 \label{fig:range_error}}
             \subfloat[Velocity error (clear)]
        {\includegraphics[width=0.33\textwidth]{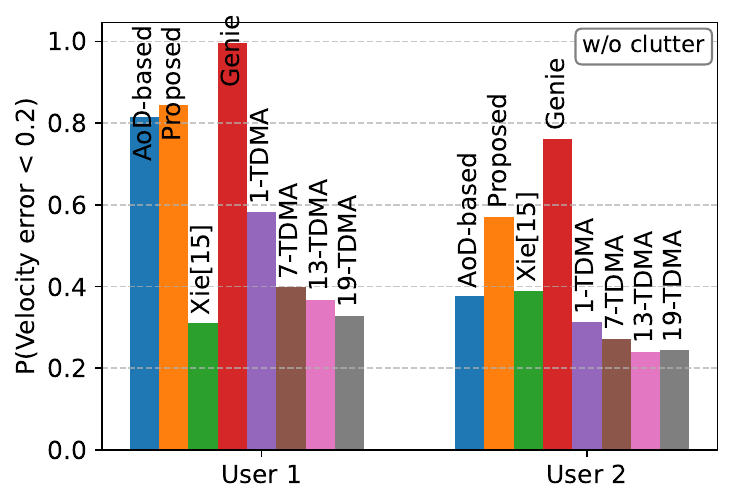}
			 \label{fig:velocity_error}}
             \subfloat[Angle error (clear)]
        {\includegraphics[width=0.33\textwidth]{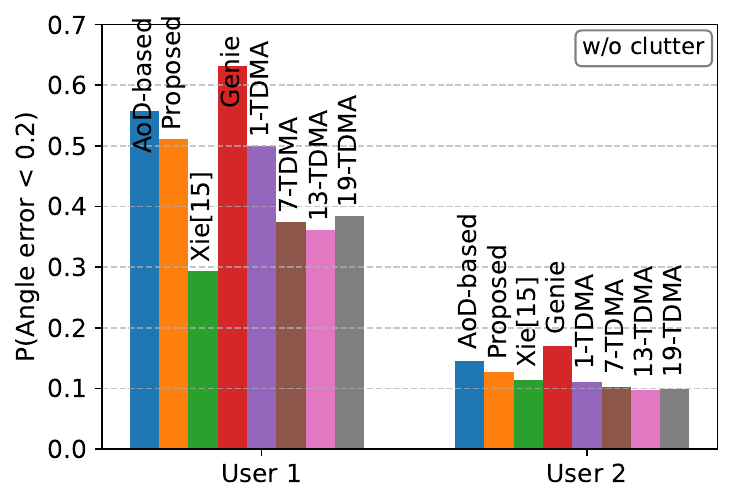}
			 \label{fig:angle_error}}
	\end{tabular}
\caption{Variation in the probabilities of estimation error less than a predefined threshold when different methods are adopted in the two channel conditions.  In (a) and (d), evaluation of range error, (b) and (e), evaluation of velocity error, and (c)(f), evaluation of angle error. In (a), (b), and (c), cluttered channel condition. In (d), (e), and (f), clear channel condition.}
\label{fig:sensing_performance}
\end{figure*}

To demonstrate the improvement provided by the proposed PPO-based method, Fig.~\ref{fig:latency} exhibits the latency of the transmitted packets in the cluttered environment. For the Genie method with user AoDs known to the transmitter, more than 70\% of the packets can be transmitted instantly. For the AoD-based method, the beams are allocated according to the AoD estimated from the previous TTI, leading to delayed beam allocation. As such, the probability of packet latency peaks at 3. For $X$-TDMA, a larger $X$ results in shorter latency. Due to a relatively slow user speed, it does not require a frequent beam sweeping to update the beams. The PPO-based method outperforms the AoD-based and $X$-TDMA methods, as the BS updates the beams properly given the observations.

In Fig.~\ref{fig:averaged_delay}, the averaged delay of PPO-based, Genie, and AoD-based methods as a function of different packet arrival probabilities is compared.  It is observed that all methods result in a higher averaged latency when there are higher packet arrival probabilities. The averaged delay of AoD-based drastically grows as the packet arrival probability increases. On the other hand, the proposed PPO-based method keeps a similar growing rate as the Genie method. This shows that the PPO policy takes buffer into consideration and minimizes the packet latency.

\subsection{Sensing Performance}
In this subsection, we evaluate the sensing performance. Fig.~\ref{fig:sensing_performance} depicts the normalized range and velocity estimation errors of two users under both cluttered and clean channel conditions. The results show that the parameter estimation performance of user 2 is worse than that of user 1, since user 2 is farther from the BS and thus has a lower SINR in the reflected echoes. Moreover, when clutter is present, the performance disparity becomes more pronounced because clutters often block the LoS path to the user.

Among the different methods, the Genie approach provides an upper bound on performance. AoD-based exhibits the most substantial performance degradation between clean and cluttered channel conditions, highlighting that it is susceptible to environmental interference in AoD estimation processes.  
For $X$-TDMA methods, a frequent multi-beam sweeping yields better sensing performance. Although PPO-based occasionally underperforms AoD-based in AoD estimation performance, it achieves enhanced range and velocity estimation accuracy. This improvement arises because the BS allocates multiple beams, which may compromise AoD estimation but provide a broader FoV to cover users.
The PPO-based method achieves performance comparable to the Genie method, especially under cluttered channel conditions. Thus, the trained PPO agent develops environmental awareness and can effectively distinguish user signals from clutter after training.

\subsection{Computational Complexity}
The proposed DRL-assisted beam management is trained offline, while online deployment only requires state constructions and policy inference at each TTI. As shown in Eq. \eqref{eq:state}, the dimension of the state is $2 N_T +2U$. Assume the policy network contains $\mathcal{L}$ fully connected layers with the respective neuron counts denoted as $n_0, n_1, \cdots, n_{\mathcal{L}}$, \textcolor{black}{where $n_0 = 2N_T + 2U$ gives the dimension of state space and $n_{\mathcal{L}} = 6^U$ denotes the dimension of action space.} Therefore, the total runtime complexity of one TTI is 
    $ \mathcal{O}\left( \sum_{l=1}^{\mathcal{L}} n_{l-1} n_{l}\right)$.
Specifically, the implemented neural network has 2 hidden layers, each with 128 nodes. Furthermore, although the number of beams $N_T$ and the number of users $U$ affect the training convergence speed, they have only a moderate impact on online computational complexity. Therefore, the DRL-assisted approach is suitable for real-time beam scheduling after the offline training.  

\subsection{Discussion}
The BS reuses the communication waveform for sensing. This ISAC scheme allows sensing without sharing many resources from communication. This can also be regarded as \emph{sensing for free}, which is an approach proposed in \cite{xie_optimal_2023}. In \cite{xie_optimal_2023}, \emph{sensing for free} occurs only when residual time resources remain after optimal scheduling is employed. In our case, sensing for free occurs continuously in the time domain by leveraging the monostatic passive radar structure. Nevertheless, the tradeoff between the two modules manifests in the number of beams allocated to users. 
Multi-beams provide a broader FoV, increasing the possibility of covering a user and thus improving the prediction of future beams. However, multiple beams spread the transmitted power, degrading communication performance. 

Traditional AoD estimation methods, such as MUSIC, achieve decent performance but suffer from user-beam association problems in multi-user environments. The association information is typically conveyed to the BS via a user feedback channel after users identify their optimal beam from BS beam sweeping. The proposed PPO-based method acquires this information based on the previous beam allocations, which are included as state variables. 

AoD estimation methods suffer performance degradation in cluttered environments due to reflections from scatterers. These additional echoes complicate the identification of user AoDs. The BS requires supplementary information to distinguish AoDs of users from those of clutters. In contrast, the PPO agent learns the environment implicitly via the neural network, adapting to distinguish user signals from interference without explicit clutter modeling. For a fair comparison, we assume the BS has user-beam association information when implementing the AoD-based benchmark method. The comparison between a clean environment and a cluttered environment is presented in Section \ref{sec:simulation}. 

In communication-only systems, determining optimal beam pairs between transmitter and receiver requires a training process in which the transmitter sends sensing packets and the receiver provides feedback. Interleaved and hierarchical search techniques are proposed to reduce this training overhead \cite{mabrouki_codebook_2022}. In ISAC systems, the sensing module potentially helps transmitters gain awareness of the wireless environment. Thus, it can reduce the training overhead for the optimal beam pair selection. As shown in Section \ref{sec:objective}, the propagation environment and beam allocation quality are reflected in the beamforming output of the echo signals. The proposed DRL-assisted beam allocation only requires acknowledgment/negative acknowledgment (ACK/NACK) and relative positions to be fed back to the BS during the training period. In addition, existing DRL-assisted resource allocation methods use communication rate and/or SNR as observation states, requiring user feedback during real-time implementation. The proposed DRL architecture, on the other hand, fully exploits the sensing module and can significantly reduce communication overhead.

\section{Conclusion}
\label{sec:conclusion}
In this paper, we investigated dynamic cross-layer beam scheduling and allocation for multi-user ISAC systems. The resource allocation problem is first formulated; it cannot be solved using traditional optimization methods because expressing communication latency under dynamic buffer conditions analytically is intractable. The proposed DRL-assisted method leverages deep neural networks to effectively capture the intricate interplay between channel dynamics and buffer states, thereby overcoming the analytical limitations of traditional approaches. We further demonstrated that monostatic ISAC can effectively leverage reflected echoes for both target sensing and beam allocation without requiring a separate sensing module or extensive user feedback. Simulation results verified that the multi-beam scheme enhances the overall throughput at the expense of modest delay increases. Moreover, the proposed DRL-assisted beam management strategy approaches the communication and sensing performance of the genie-aided benchmark with perfect AoD knowledge.  These findings collectively provide insights for the future design of learning-based resource management in ISAC systems.

\bibliographystyle{IEEEtran}
% \IEEEtriggeratref{8}
% \bibliography{centric}

\end{document}